\documentclass{aa}  
\usepackage{graphicx}
\usepackage{adjustbox}
\usepackage{amsmath}
\usepackage{color}
\usepackage{txfonts}
\usepackage{hyperref}
\usepackage{amsmath}
\usepackage{xcolor}

\usepackage{media9}

\usepackage{caption}

\begin{document}

\title{Poynting flux transport channels formed in polar cap regions of neutron star magnetospheres}

\titlerunning{Poynting flux channels formed in pulsar polar caps}

\author{Jan Ben\'{a}\v{c}ek \inst{1},
      Andrey Timokhin \inst{2},
      Patricio~A.~Mu\~noz \inst{4,3},
      Axel~Jessner\inst{5}, \\
      Tatiana Rievajová \inst{6},
      Martin Pohl \inst{1,7},
      and Jörg Büchner \inst{4,3}
      }
      
\authorrunning{Benáček et al.}

\institute{
    Institute for Physics and Astronomy, University of Potsdam, 14476 Potsdam, Germany \\ \email{jan.benacek@uni-potsdam.de}
    \and
    Janusz Gil Institute of Astronomy, University of Zielona Góra, ul. Szafrana 2, 65516, Zielona Góra, Poland
    \and
    Max Planck Institute for Solar System Research, 37077 G\"ottingen, Germany
    \and
    Center for Astronomy and Astrophysics, Technical University of Berlin, 10623 Berlin,
    Germany 
    \and
    Max Planck Institute for Radio Astronomy, 53121 Bonn, Germany
    \and 
    Institute of Theoretical Physics and Astrophysics, Masaryk University, 611 37 Brno, The Czech Republic
    \and 
    Deutsches Elektronen-Synchrotron DESY, Platanenallee 6, D-15738 Zeuthen, Germany
}

\date{Received: 31 May 2024; accepted: 15 September 2024}

% \abstract{}{}{}{}{} 
% 5 {} token are mandatory
 
  \abstract
   { Pair cascades in polar cap regions of neutron stars are considered to be an essential process in various models of coherent radio emissions of pulsars. 
       The cascades produce pair plasma bunch discharges in quasi-periodic spark events.
       The cascade properties, and therefore also the coherent radiation, depend strongly on the magnetospheric plasma properties and vary significantly across and along the polar cap.
       Importantly, where the radio emission emanates from in the polar cap region is still uncertain.
   }
   {
       We investigate the generation of electromagnetic waves by pair cascades and their propagation in the polar cap for three representative inclination angles of a magnetic dipole, $0^\circ$, $45^\circ$, and $90^\circ$.
   }
   {
       We use two-dimensional particle-in-cell simulations that include quantum-electrodynamic pair cascades in a charge-limited flow from the star surface.
   }
   {
       We find that the discharge properties are strongly dependent on the magnetospheric current profile in the polar cap and that 
    transport channels for high intensity Poynting flux are formed along magnetic field lines where the magnetospheric currents approach zero and where the plasma cannot carry the magnetospheric currents.
    There, the parallel Poynting flux component is efficiently transported away from the star and may eventually escape the magnetosphere as coherent radio waves.
    The Poynting flux decreases with increasing distance from the star in regions of high magnetospheric currents.
   }
   {
       Our model shows that no process of energy conversion from particles to waves is necessary for the coherent radio wave emission.
       Moreover, the pulsar radio beam does not have a cone structure; rather, the radiation generated by the oscillating electric gap fields directly escapes along open magnetic field lines in which no pair creation occurs.
   }


   \maketitle
%
%-------------------------------------------------------------------

%%%%%%%%%%%%%%%%%%%%%%%%%%%%%%%%%%%%%%%%%%%%%%%%%%%%%%%%%%%%%%%
%%%%%%%%%%%%%%%%%%%%%%%%%%%%%%%%%%%%%%%%%%%%%%%%%%%%%%%%%%%%%%%
%%%%%%%%%%%%%%%%%%%%%%%%%%%%%%%%%%%%%%%%%%
\section{Introduction} \label{sec:intro}
It is widely assumed that the coherent radio emission in rotation-powered pulsars originates in or close to their polar cap regions where pair cascades create a dense electron-positron plasma \citep{Sturrock1971,Ruderman1975,Cheng1977b,Daugherty1982}. 
However, the exact coherent radio emission mechanism and its location remain uncertain \citep{Melrose2017a,Melrose2020a,Philippov2022}.

The polar cap pair cascades play an essential role in filling the neutron star magnetosphere with plasma \citep{Tomczak2023}, powering the plasma instabilities behind coherent radio emissions \citep{Asseo1998,Melikidze2000,Gil2004}, as well as in powering the pulsar wind \citep{Petri2022}, heating the neutron star surface \citep{Zhang2000,Gonzales2010,Kopp2023}, and in the emission of a wide spectral range of electromagnetic waves \citep{Hobs2004,Daugherty1996,Petri2019,Giraud2021,Benacek2024,Labaj2024}.
Dense plasma trapped in the closed magnetic field line zone of the pulsar magnetosphere screens the electric field component parallel to the magnetic field, thus suppressing particle acceleration in this zone. 
There is no poloidal current flowing along closed magnetic field lines, and the plasma co-rotates with the neutron star. 
Along open magnetic field lines, the plasma flows away from the neutron star, forming the pulsar wind. 
The flow of this plasma generates poloidal currents supporting the force-free structure of the pulsar magnetosphere \citep{Goldreich1969}. 
The plasma has to be constantly replenished, as it is leaving the inner magnetosphere. 
Most of this plasma is generated in electron-positron cascades in pulsar polar caps --- regions close to the magnetic poles --- as first suggested by \citet{Sturrock1971}. 
These cascades are initiated by ultrarelativistic particles accelerated in the polar cap regions with strong parallel electric fields called gaps. 
The electric field appears when the plasma cannot sustain the poloidal current density and cannot provide the charge density required to maintain the twist of open magnetic field lines and the corresponding electric fields so that the magnetosphere remains force-free \citep{Beloborodov2008,Timokhin2010,Timokhin2013}.
Particles accelerated to high energies in the gaps emit $\gamma$-ray photons that interact with the super strong magnetic field, thereby creating electron--positron pairs.

Pair creation occurs in the form of repetitive cascade events.
When a sufficient amount of particles has been produced in a cascade, the motion of the plasma charges begins to support the required poloidal currents, and the charge separation can sustain the required charge density. 
 This causes the screening of the electric field and the cessation of pair creation.
When the newly created plasma leaves the polar cap, the plasma density drops again, the necessary current and charge densities cannot be sustained, and the discharge repeats.
The cascade events in polar caps were first studied through particle-in-cell (PIC) 1D (one dimension in space) electrostatic simulations from first principles by \citet{Timokhin2010,Timokhin2013}.
The $\gamma$-ray photons in those simulations were produced by curvature radiation of the curved particle motion in the magnetosphere.
It was shown that the value of the required poloidal current density strongly influences the discharge behavior and their repetition --- the burst of pair creation is followed by a longer phase when the dense freshly created plasma flows into the magnetosphere, and the accelerating field is fully screened. 
The total pair production efficiency of polar cap cascades has recently been studied in \citet{Timokhin2015, Timokhin2019}, and it was shown that the number of produced particles does not exceed approximately a few$\times$10$^{5}$ per primary particle.
\citet{Cruz2021a,Cruz2022} investigated the filling of the magnetosphere by pairs using the PIC simulations, the generation of large-amplitude oscillating electrostatic waves, and the growth rates and screening times of the magnetospheric electric fields.
\citet{Okawa2024arXiv} studied pair creation and parallel electric field screening.
They showed that the time-averaged power spectrum is similar to a typical pulsar radio spectrum.
The $\gamma-$photon polarization during the QED cascade was considered by \citet{Song2024arXiv}, and it was revealed that during the cascade, the synchrotron losses may dominate over those from curvature radiation.

The 1D electrostatic simulations of \citet{Timokhin2010,Timokhin2013} showed that the intermittent screening of the electric field during pair discharges generates strong superluminal electrostatic waves. 
It was speculated that in multi-dimensional discharges, these waves might become electromagnetic waves, with a component of the electric field perpendicular to the magnetic field $\boldsymbol{B}$. 
If these waves escape the plasma, they can become observable as coherent radio emission. 
\citet{Philippov2020} further developed this idea and proposed a novel mechanism for the generation of electromagnetic waves in the plasma during pair discharges. 
In this mechanism, the inhomogeneity of pair production across the magnetic field lines causes the appearance of a perpendicular (to $\boldsymbol{B}$) component of the electric field, and thus, despite all particles in the discharge moving strictly along the magnetic field lines, the discharge generates an electromagnetic mode.
They performed the first 2D simulations of pair cascades and confirmed the appearance of such waves during pair discharges. 
They showed that the spectra of such waves are broadband, and their frequency ranges are commensurate with the observed pulsar spectra. 
\citet{Tolman2022} studied the damping of such waves during discharges when plasma density grows rapidly due to ongoing pair creation. 
It was shown that after initially strong exponential damping, the damping of the waves becomes linear in time.
This indicates that these waves could still contain enough energy when they decouple from plasma and become electromagnetic waves observed as pulsar radio emission when plasma density drops at some distance from the discharge zone.
\citet{Benacek2023} studied the linear acceleration emission of the time-evolving plasma bunches through postprocessing of results from PIC simulations and found that the emission spectrum is comparable with the observed one for typical pulsars.

The radio emission mechanism suggested by \citet{Philippov2020} has recently been studied by other authors as well.
\citet{Cruz2021b} studied the polar cap of an aligned pulsar using a more realistic setup than in \citet{Philippov2020} by considering cascades along diverging magnetic field lines. 
In their simulations, they observed two peaks in the outflowing Poynting flux of electromagnetic plasma waves that could be associated with core and conal components of pulsar radio emissions \citep{Rankin1990,Rankin1993}.
\citet{Bransgrove2023} performed global 2D PIC simulations of the aligned rotator's magnetosphere and demonstrated the excitation of electromagnetic waves in pair discharges whenever they occur in the magnetosphere.

However, the details of how the pair cascades produce the observed radiation are still lacking.
In addition, how the generated Poynting flux and hence the properties of the produced radio waves are influenced by the local kinetic plasma properties and the global magnetospheric environment is an open question.

In this paper, we focus on how the pulsar inclination angle, the angle between the pulsar's magnetic moment and its angular velocity of rotation, and thus the magnetospheric current profile across the polar cap influences the discharges, the production of plasma bunches, and the Poynting flux of electromagnetic waves that are generated in the discharges.
We investigate the time evolution of pair cascades in the polar cap region under the assumption of charge limited flow, where particles can freely leave the neutron star surface \citep{Arons1979} using
kinetic and relativistic PIC simulations from first principles. 
In our model, the discharges are driven by the magnetospheric currents and their profile across the polar cap, the latter being known from global magnetospheric general-relativistic force-free simulations that include a return current.

This paper is structured as follows.
In Sect.~\ref{sec:methods} we discuss the plasma and numerical setup of the particle in cell simulations.
Section~\ref{sec:results} describes the cascade evolution and shows the properties of the cascade events after reaching a quasi-periodic state.
We discuss the cascade and radiation properties in Sect.~\ref{sec:discuss} and state the main conclusions about the Poynting flux properties in Sect.~\ref{sec:conclude}.

%%%%%%%%%%%%%%%%%%%%%%%%%%%%%%%%%%%%%%%%%%%%%%%%%%%%%%%%%%%%%%%%%%%%%%%%%%%%%%%%%%%%%%%%%%%%
\section{Methods}  \label{sec:methods}

\subsection{Polar cap model}

\begin{figure}[t]
    \includegraphics[width=0.5\textwidth]{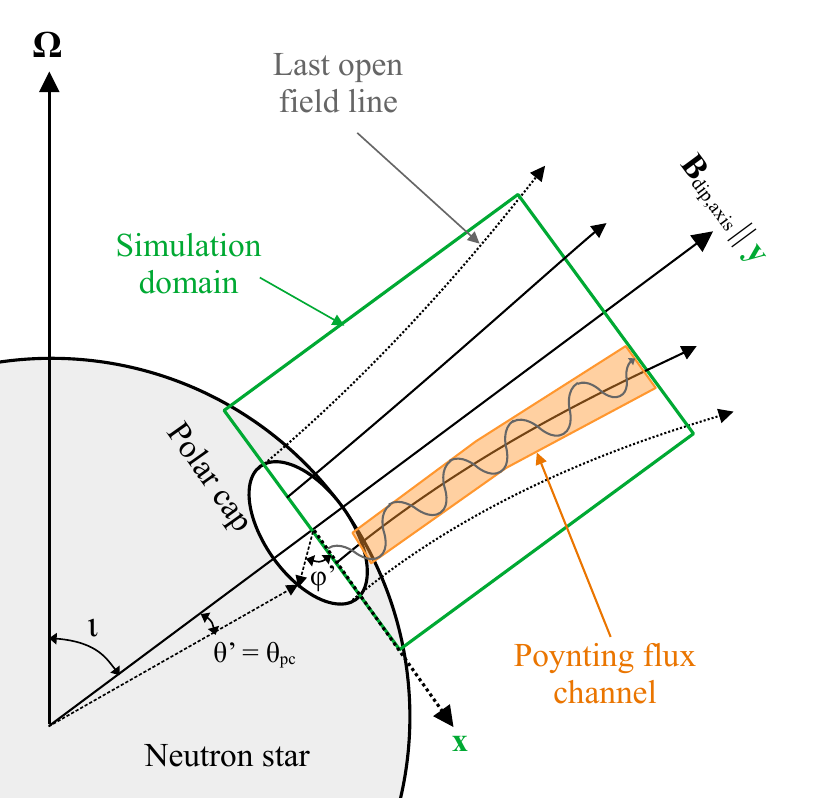}
    \caption{
        Scheme of the neutron star polar cap in the considered model.
        Not to scale.
        \label{fig0}
    }
\end{figure}

% Simulation parameters
\begin{table*}[htp]
        \centering
    \caption{Parameters for the star and simulation.
    }
        \begin{tabular}{lccc}
                \hline \hline
        Parameter & \multicolumn3c{Values} \\
        \hline
        Dipole inclination $\iota$ & 0$^\circ$ & 45$^\circ$ & 90$^\circ$ \\
        Dipole magnetic field $B_\mathrm{dip,axis}$ & \multicolumn3c{10$^{12}$\,G} \\
        Star rotation period $P$ & \multicolumn3c{0.25\,s} \\
        Star mass $M_\star$ & \multicolumn3c{1.5\,M$_\mathrm{s}$} \\
        Star radius $R_\star$ & \multicolumn3c{12\,km} \\
        \hline
        Domain size $L_x$ & 11\,000$\Delta_\mathrm{x}$ & 11\,000$\Delta_\mathrm{x}$ & 9\,000$\Delta_\mathrm{x}$ \\
        Domain size $L_y$ & \multicolumn3c{12000$\Delta_\mathrm{x}$} \\
        Grid cell size $\Delta_\mathrm{x}$ & \multicolumn3c{13.4\,cm} \\
        Initial plasma skin depth $d_\mathrm{e}(n_0) / \Delta_\mathrm{x}$ & \multicolumn3c{14.93} \\
        Simulation time step $\Delta t$ & \multicolumn3c{0.2011\,ns} \\
        Simulation time steps $T / \Delta t$ & \multicolumn3c{150\,000} \\
        Polar cap angle $\theta_\mathrm{pc}$, Eq.~\ref{eq:cap_angle} & 2.01$^\circ$ & 2.15$^\circ$ & 2.20$^\circ$ \\
        Polar cap transition angle $\Delta\theta / \theta_\mathrm{pc}$ & \multicolumn3c{0.01} \\
        Current profile $j_\parallel / j_\mathrm{GJ}$ & \multicolumn3c{Eq.~\ref{eq:current_profile}, Fig.~\ref{fig1}} \\
        Particle decay threshold $\gamma_\mathrm{thr}$ & \multicolumn3c{$2\times 10^{7}$} \\ 
        Secondary particle $\gamma_\mathrm{sec}$ & \multicolumn3c{$1 \times 10^{4}$}\\ 
        Friedman filter $\theta_\mathrm{Fr}$ & \multicolumn3c{0.1} \\
        \hline
        Initial macro-particle density representing $n_0$ & \multicolumn3c{2\,PPC} \\
        Initial and injection particle velocity $u_\mathrm{0} / c$ & \multicolumn3c{0.1} \\
        Initial density closed lines $n_\mathrm{closed}$ & \multicolumn3c{$50\,n_0$} \\
        Absorbing boundary condition width $l_\mathrm{abc}$ & \multicolumn3c{40$\Delta_\mathrm{x}$} \\
                \hline \hline
        \end{tabular}
    \\ \footnotesize{}
        \label{tab1}
\end{table*}

% Star parameters + magnetic field
We assume a spherical neutron star with a mass $M_\star = 1.5\,M_\mathrm{s}$, where $M_\mathrm{s}$ is the solar mass, a radius $R = 12$\,km, and a rotation period $P = 0.25$\,s.
We denote the dipole magnetic field as $\boldsymbol{B}_\mathrm{dip}$ and the magnetic field in the dipole axis at the star surface as $\boldsymbol{B}_\mathrm{dip,axis} = \boldsymbol{B}_\mathrm{dip}(\iota, R_\star)$.
The dipole axis has an inclination angle $\iota$ to the star rotational axis $\boldsymbol{\Omega}$.
We study three dipole inclinations of $\iota = 0^{\circ}, 45^\circ,$ and $90^{\circ}$.
The summary of all simulation parameters is in Table~\ref{tab1}.

% Polar cap properties
The polar cap has a radius that is defined by the "last open magnetic field line" \citep{Gralla2017}:
\begin{equation} \label{eq:cap_angle}
    \alpha_\mathrm{pc}(\iota) = \sqrt{\frac{3}{2}} \mu \Omega \left( 1 + \frac{1}{5} \sin^2 \iota \right),
\end{equation}
where $\Omega = 2\pi / P$ is the star angular velocity and $\mu$ is the dipole magnetic moment.
For the considered inclinations $\iota$, the last open magnetic field lines correspond to polar cap angles:
    \[
    \theta_\mathrm{pc}(\iota) \equiv \arcsin  \sqrt{\frac{\alpha_\mathrm{pc}R_\star}{\mu}} \approx 2.01^{\circ}, 2.15^{\circ}, \mathrm{and} \, 2.20^{\circ}.
\]
The polar cap is described in spherical coordinates $(\theta,\varphi, r)$ centered at the star center. 
The polar angle $\theta$ is measured from the magnetic dipole axis, and the azimuthal angle $\varphi$ is denoted as in Fig.~\ref{fig0}.
In addition, we assume a transition angle $\theta \in (\theta_\mathrm{pc},\theta_\mathrm{pc} + \Delta \theta)$ between the closed and open field lines, with $\Delta \theta = 0.01 \, \theta_\mathrm{pc}$.

The magnetic field is constructed as a composition of two fields:
\begin{equation}
    \boldsymbol{B}(\boldsymbol{x},t) = \boldsymbol{B}_\mathrm{dip}(\boldsymbol{x}) + \delta \boldsymbol{B}(\boldsymbol{x},t),
\end{equation}
where $\delta \boldsymbol{B}$ is the local space- and time-varying field, and $\boldsymbol{B}_\mathrm{dip}$ is the external dipole field that does not change its structure in the polar cap in time.
Close to the star, this approximation is valid as the light cylinder distance is $P c/2 \pi \approx 10^3 R_\star $.

\subsection{Particle-in-cell simulations}
% PIC model
Because global kinetic magnetospheric simulations on current supercomputers do not resolve the kinetic scales of a typical pulsar polar cap, we investigate the polar cap in a local frame co-rotating with the neutron star.
The plasma is described at kinetic scales using a fully kinetic, relativistic, and electromagnetic particle-in-cell (PIC) code. 
We utilize the 2D3V version of the code ACRONYM with a rectangular grid \citep{Kilian2012}.

% ACRONYM
We use the fourth order M24 finite-difference time-domain (FDTD) method proposed by \citet{Greenwood2004} for computation on the Yee lattice to efficiently describe the wave dispersion properties of relativistic plasmas.
We combine the field solver with the \citet{Friedman1990} low-pass filtering method with the Friedman parameter $\theta_\mathrm{Fr} = 0.1$ to weakly suppress the numerical Cerenkov radiation.
For the current deposition, we utilize the \citet{Esikperov2001} current-conserving deposition scheme with a fourth-order ``weighting with time-step dependency'' (WT4) shape function of macro-particles \citep{Lu2020}.
We use the \citet{Vay2011} particle pusher that was modified to advance particles in the gyro-motion approximation \citep{Philippov2020}.
The approximation assumes a strict particle motion along the direction of the local magnetic field vector, and the pusher therefore utilizes only the parallel component of the electric field.
The approach is valid because the particles are assumed to lose their kinetic energy perpendicular to the magnetic field in a much shorter time than one simulation time step.
The particle pusher assumes that the gravitational force of the neutron star has only a weak impact on non-relativistic particles.

% Numerical domain + scaling
The simulation uses a uniform rectangular grid of size $L_\mathrm{x} \times L_\mathrm{y}$.
(For a scheme, please see Fig.~\ref{fig0}.)
The length $L_\mathrm{y}$ is fixed to $12\,000 \Delta_\mathrm{x}$ along the dipole axis, where $\Delta_\mathrm{x} = 13.4$\,cm is the grid size.
The length $L_\mathrm{x}$ is varied to cover the whole polar cap angle with the transition angle and the return current, giving $L_\mathrm{x} = 11\,000\Delta_\mathrm{x}, 11\,000\Delta_\mathrm{x},$ and $9\,000\Delta_\mathrm{x}$ for $\iota = 0^{\circ}, 45^\circ,$ and $90^{\circ}$, respectively.
For example, for $\iota = 0^{\circ}$, the real size of the domain is 1474\,m $\times$ 1608\,m.
The simulation length $L_\mathrm{y}$ covers larger distances than the typically assumed electric gap size of $l_\mathrm{gap} \simeq 50-100$\,m, and maintains the scaling relations $L_\mathrm{y} \gg l_\mathrm{gap} \gg \Delta_\mathrm{x}$.
The time step is chosen as $\Delta t = 0.45 \Delta_\mathrm{x}/c \approx 0.2011$\,ns.

% Simulation orientation
The simulation domain is a 2D cut through the 3D polar cap in the $\boldsymbol{\Omega} - \boldsymbol{B}_\mathrm{dip,axis}$ plane.
In all three cases, the start of the coordinate system ($x=0$, $y=0$) is at the dipole axis below the star surface ($r \lesssim R_\star$).
% Also, the simulation domain covers open and closed magnetic field lines.
The star surface is located $\sim 70 \Delta_\mathrm{x} \approx 9.4$\,m from the simulation boundary at the dipole axis. 
As the surface bends, it approaches the simulation boundary to 20 grid cells, $\approx 2.7$\,m, at the last open field line.
The simulation $y$-axis is always parallel to $\boldsymbol{B}_\mathrm{dip,axis}$.

\subsection{Magnetospheric currents}
% Magnetospheric currents
We added global magnetospheric currents $j_\mathrm{mag}(\boldsymbol{x})$ by modifying the PIC field solver as \citet[Appendix~A]{Timokhin2010}
\begin{equation} \label{eq:electric_field}
    \frac{\partial \boldsymbol{E}(\boldsymbol{x},t)}{\partial t} = - 4\pi\left(\boldsymbol{j}(\boldsymbol{x},t) - \boldsymbol{j}_\mathrm{mag}(\boldsymbol{x}) \right) + c (\nabla \times \delta\boldsymbol{B}(\boldsymbol{x},t)).
\end{equation}
The term $(\nabla \times \delta\boldsymbol{B})$ only contains the time-evolving magnetic field.
The current density $j(\boldsymbol{x},t)$ is caused by the plasma motion.
The term $\boldsymbol{j}_\mathrm{mag}(\boldsymbol{x})$ in Eq.~\ref{eq:electric_field} represents the current density ``demanded'' by the magnetosphere --- the current density necessary to support the twist of magnetic field lines in the force-free magnetosphere.

We assumed the current $\boldsymbol{j}_\mathrm{mag}$ is always parallel to the dipole magnetic field:
\begin{equation} \label{eq:magnetospheric_current}
    \boldsymbol{j}_\mathrm{mag}(\boldsymbol{x}) = j_\parallel(\boldsymbol{x}) \frac{\boldsymbol{B}_\mathrm{dip}(\boldsymbol{x})}{|\boldsymbol{B}_\mathrm{dip}(\boldsymbol{x})|}.
\end{equation}
We express the magnetospheric parallel current $j_\parallel$ separately for the open and closed magnetic field lines.
In open field lines, it is expressed as an analytical fit of the currents across the polar cap that was found by \citet{Gralla2017} and \citet{Lockhart2019} in general-relativistic force-free simulations:
\begin{equation} \label{eq:current_profile}
    \boldsymbol{j}_\mathrm{mag} = \frac{\Lambda}{\sqrt{\Upsilon}} \boldsymbol{B}, \\
\end{equation}
\begin{multline}
    \Lambda = \mp 2 \Omega \left\{ J_0\left(2 \arcsin \sqrt{\frac{\alpha}{\alpha_\mathrm{pc}}}\right) \cos \iota  \right. \\
    \mp \left. J_1\left(2 \arcsin \sqrt{\frac{\alpha}{\alpha_\mathrm{pc}}}\right) \cos \beta \sin \iota \right\}, \qquad
    \alpha < \alpha_0,
\end{multline}
\begin{equation}
    \Upsilon \approx 1 - \frac{2 G M_\star}{c^2} \frac{1}{r},
\end{equation}
where $J_0$ and $J_1$ are the Bessel functions of the first kind, 
$\Upsilon$ is the reddening factor, $G$ is the gravitational constant, and $M_\star$ is the star mass,
The "Euler potentials" of the magnetic dipole $(\alpha, \beta)$ are represented in spherical coordinates $(\theta ', \varphi ', r)$ of the polar cap region,
\begin{equation}
    \alpha = \frac{\mu}{r} \sin^2 \theta ', \qquad \beta = \varphi ',
\end{equation}
and refer to the dipole axis. 
The $\mp$ signs in brackets of Eq.~\ref{eq:current_profile} correspond to the north and south magnetic poles, respectively.

The \citet[GJ]{Goldreich1969} charge density,
\begin{equation} \label{eq:GJcharge}
    \rho_\mathrm{GJ} = -\frac{\boldsymbol{\Omega} \cdot \boldsymbol{B}_\mathrm{dip}(\boldsymbol{x},\iota)}{2\pi c},
\end{equation}
is generally a function of the dipole angle and the position in the magnetosphere. 
For an anti-aligned rotator $\Omega < 0$ that we consider in our case, it follows from Eq.~\ref{eq:GJcharge} that the studied polar cap region always has positive GJ charge density for dipole inclinations $0^\circ$ and $45^\circ$.
The GJ charge density is positive and negative for $x<0$ and $x>0$, respectively, for the inclination $90^\circ$.

The local GJ current for an aligned pulsar is
\begin{equation}
    j_\mathrm{GJ} \equiv c \rho_\mathrm{GJ}\left(B_\mathrm{dip}(\boldsymbol{x},\iota)\right) \equiv - \frac{\Omega B_\mathrm{dip}(\boldsymbol{x},\iota)}{2 \pi},
\end{equation}
with a specific case of GJ current in the dipole axis at star surface for an aligned pulsar,
\begin{equation}
    j_\mathrm{GJ,axis} \equiv j_\mathrm{GJ}(\boldsymbol{x}=\boldsymbol{0}, \iota=0).
\end{equation}

Our model incorporates a return current so that the net magnetospheric current through the simulation boundary vanishes.
The return current for inclinations $\iota = 0^\circ$ and $\iota = 45^\circ$ was modeled to have a half-sine profile as a function of the angle $\theta'$ in the closed field lines. The return current amplitude was fixed to $j_\mathrm{mag}\sqrt{\Upsilon}/j_\mathrm{GJ,axis} = -6$ and the width of the return current profile in closed field lines was adjusted so that the resulting net current is zero at the star surface for each simulation.
If the magnetospheric current Eq.~\ref{eq:current_profile} is non-zero on the last open field line, a part of the sinusoidal wave was truncated, thereby maintaining the smooth functional form of the current profile between open and closed field lines.
The net current is zero for $\iota = 90^{\circ}$; no return current was added.

% Figure 1 - current profile across polar cap
\begin{figure}[t]
    \includegraphics[width=0.45\textwidth]{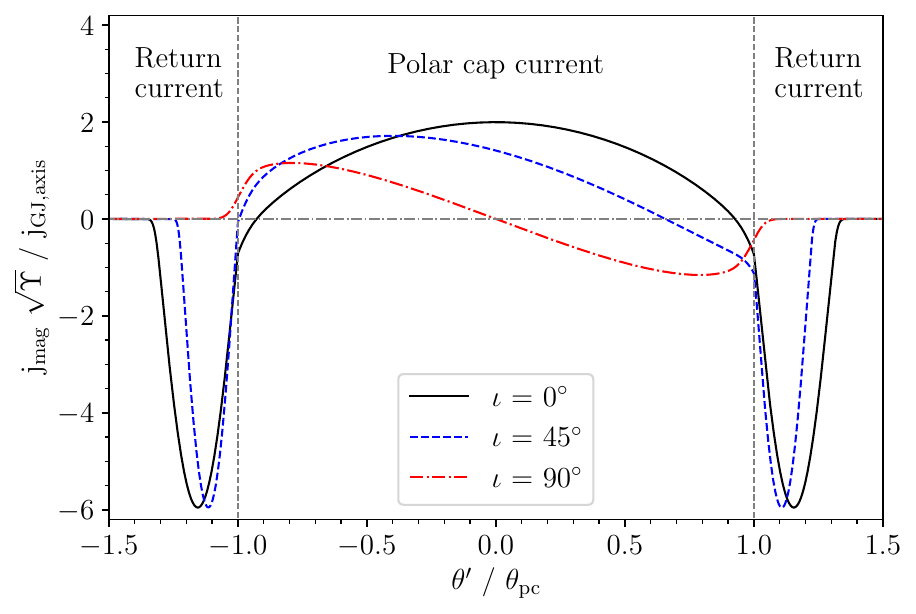}
    \caption{
        Magnetospheric current profile across the polar cap at the star surface for the dipole inclination angles we used.
        The currents have been normalized to the GJ current density in the dipole axis at the star surface for an aligned pulsar, $j_\mathrm{GJ,axis}$, and the reddening factor, $\Upsilon$.
        \label{fig1}
    }
\end{figure}
The simulation plane is a cut across the polar cap.
We define the cut for $\varphi' = 0$ as $\theta' \ge 0$ and for $\varphi' = \pi$ as $\theta < 0$.
We smooth the magnetospheric current profiles by convolution with a Gaussian function with a width of $0.04\theta_\mathrm{pc}$ in order to obtain a smooth transition between open and closed field lines.
The resulting global magnetospheric current profiles used in simulations are shown in Fig.~\ref{fig1}.

\subsection{Boundary and initial conditions}
% Initial conditions
We assume that the initial electric field is zero in the simulation domain and that the magnetic field $B_\mathrm{ext}$ is given only by the dipole magnetic field.
The open magnetic field lines are initially filled with $n_0 = 2$ macro-particles per cell (PPC), and the closed field lines by $n_\mathrm{closed} = 10$\,PPC, half electrons and half positrons.
The macro-particles in closed field lines represent ten times more plasma particles than those in open field lines; that is, their physical density is 50 times higher than in closed field lines.
Therefore, not many macro particles are required to sustain the relatively large return current in closed field lines.
The plasma in open magnetic field lines is not dense enough to fully sustain the magnetospheric currents.
We implemented a linear transition in density between these regions of the angular width $\Delta \theta$.
That corresponds to $\approx$$34 \Delta_\mathrm{x} \approx 4.5$\,m on the star surface.
This width assures that the transition between both regions is not too steep for an appropriate representation by the simulation grid.
% that would  otherwise only be poorly described by the simulation grid.

When the simulation starts, all particles have the Maxwell-Boltzmann distribution function with a thermal velocity of $u_\mathrm{0} = 0.1\,c$.
The particles are then accelerated by electric fields.
The particles from the initial filling of the simulation domain are not accelerated in the first time step because the electric fields are initially set to zero everywhere; however, they may be accelerated in the following time steps when the magnitude of the electric field increases as the result of the magnetospheric currents described by Eq.~\ref{eq:electric_field}.
We tested various thermal velocities, but they appear to have only a negligible impact on the simulation results as long as the Lorentz factor $\gamma(u_\mathrm{0}) \ll \gamma_\mathrm{thr}'$, where the $\gamma_\mathrm{thr}'$ is the threshold Lorentz factor defined below.

% Boundary conditions for particles
The simulations have open boundary conditions for particles.
We inject particles into the simulation in every time step, but differently for open and closed magnetic field lines.
For the open field lines, we inject particles on the star ``surface'' and at the ``top'' boundary far from the star ($y = L_\mathrm{y}$) if the number density drops below a threshold of $n_0 = 2$\,PPC $\approx 2n_\mathrm{GJ,axis}$, where $n_\mathrm{GJ,axis} \equiv \rho_\mathrm{GJ}(B_\mathrm{dip,axis}) / e$.
We also tested higher injection rates at the star surface; however, there is no significant difference when the system reaches the quasi-periodic pair creation stage.
The injection is into a layer of thickness of one grid cell.
The grid cells of closed magnetic field lines are filled in when the macro-particle number density in the cell drops below a threshold of $n_\mathrm{closed}$.
Nonetheless, the injection in closed field lines eventually leads to an average number density that is higher than $n_\mathrm{closed}$.
The density $n_\mathrm{closed}$ is only a minimal threshold, but particle motion allows for fluctuations above this threshold.

% Properties of injection
New particles are always injected as an electron--positron pair in order to maintain the charge neutrality of the injection.
Pairs are also injected on the surface of the star, similar to \citep{Cruz2021b}.
The positrons injected at the star surface may, in principle, become seeds of the cascade.
If they move toward the surface where they are absorbed, their contribution to the gap formation is negligible. 
The same would be true for ions. 
If ions are injected and move outward, they would not be able to produce curvature radiation capable of pair production under these circumstances.
However, inflowing positrons may serve as primary particles for the cascade, as we explain in detail in Sect.~\ref{sect:3.4}.

The injected particles have the same initial distribution as the initial particle distribution for the first step.
If there is a non-zero parallel electric field, the particles are accelerated in opposite directions after the injection because of their opposite charges.
Typically, for the injection in open magnetic field lines, one particle is removed in a few time steps when it reaches the simulation boundary, but the other one may move along the field line across the simulation domain.
The injection in the open field lines effectively eliminates all plasma cavities with zero PPC.
No cascade could occur in such cavities even if the electric fields are strong enough to accelerate particles to the cascade energies.
Thus, the few injected particles may serve as the seed particles for the cascade.

% Electric fields below star surface
To keep the electric fields under the star surface realistically close to zero, we maintain a density of $n_\mathrm{closed}$ also there, similarly as for closed magnetic field lines.
This way, the electric fields under the star surface and in closed field lines become much smaller than those in open field lines.

% Boundary conditions for field
We utilize the absorbing boundary condition in the ghost cells surrounding the simulation domain provided by the Complex Shifted Coefficient --- Convolutionary Perfectly Matched Layer (CFS---CMPL) algorithm \citep[Chapter~7]{Roden2000,Taflove2005} for the boundary conditions of electromagnetic waves.
We found that the algorithm was causing numerical artifacts when magnetospheric currents were introduced as a modification of Eq.~\ref{eq:electric_field}.
The artifacts appeared as strong electromagnetic waves propagating at an oblique angle from the last closed magnetic field line that passes through the simulation boundary.
To suppress the numerically generated waves, we added a new layer of absorbing boundary suppressing the perpendicular (but not parallel) wave oscillations around the simulation domain of thickness 40 grid cells and in all cells in the region of the closed magnetic field line.
The suppression coefficient has a linear transition that decreases with the distance from the boundaries inward of the domain.
We found this solution to be stable, not impacting the events in the open magnetic field lines.

\subsection{Scaling of plasma quantities}

% Simulation parameters
\begin{table}[tp]
        \centering
\caption{Scaling of the physical quantities in simulations.}
        \begin{tabular}{ll}
                \hline \hline
        Parameter Name & Simulation Scaling \\
        \hline
        Time & $t' = t$ \\
        Position vector & $\boldsymbol{x}' = \boldsymbol{x}$ \\
        Light speed & $c' = c$ \\
        \hline
        Plasma density & $n_\pm' = n_\pm / \zeta$ \\
        Plasma frequency & $\omega_\mathrm{p}' =  \omega_\mathrm{p} / \zeta^{\frac{1}{2}}$ \\
        Plasma skin depth & $d_\mathrm{e}' = d_\mathrm{e}\zeta^{\frac{1}{2}}$ \\
        Electric current density & $j' = j / \zeta$ \\
        Electric field intensity & $E' = E / \zeta$ \\
        Magnetic field intensity & $B' = B/\zeta $ \\
        Poynting flux & $S' = S / \zeta^2$ \\
        Threshold Lorentz factor & $\gamma_\mathrm{thr}' = \gamma_\mathrm{thr} / \zeta$ \\
        Secondary  particle Lorentz factor & $\gamma_\mathrm{sec}' = \gamma_\mathrm{sec} / \sqrt{\zeta}$ \\
                \hline \hline
        \end{tabular}
    \\ \footnotesize{The scaling factor is $\zeta$. The simulation quantities are denoted by primes.}
        \label{tab2}
\end{table}

% Scaling, time step, plasma frequency
Because the simulation grid size does not resolve the electron skin depth of the real plasma in the polar cap, the plasma parameters associated with the skin depth are scaled by a factor $\zeta$:
\begin{equation} \label{eq:scaling}
    \zeta \equiv \left(\frac{d_\mathrm{e,real}}{d_\mathrm{e,simulation}'}\right)^2 = \frac{j_\mathrm{real}}{j_\mathrm{simulation}'} = 10^4.
\end{equation}
The initial plasma density $n_0 \approx 5.6\times 10^{9}$\,cm$^{-3}$ resolves the skin depth by about 7$\Delta_\mathrm{x}$ and corresponds to the simulation plasma frequency of $\omega_\mathrm{p0}' = \sqrt{18} \times 10^{9}$\,s$^{-1}$.
Hence, the real plasma frequency is
    \begin{multline}
        \omega_\mathrm{p0,real} = \left(\frac{d_\mathrm{e,real}}{d_\mathrm{e,simulation}}\right) \omega_\mathrm{p0}
                           =  \sqrt{\zeta}\, \omega_\mathrm{p0} 
                           = \sqrt{18} \times 10^{11}\,\mathrm{s}^{-1},
    \end{multline}
and the plasma density is $n_\mathrm{real} \approx 5.6\times 10^{13}$\,cm$^{-3}$.
When the plasma density reaches a local maximum in a bunch, the skin depth is resolved by $\gtrsim$2$\Delta_\mathrm{x}$.
The mean Lorentz factor of particles will significantly decrease the plasma frequency and consequently increase the resolution of the skin depth.
The scaling of Eq.~\ref{eq:scaling} also applies to the magnetospheric currents and electromagnetic fields because the scaled magnetospheric currents influence the electric and magnetic fields in Eq.~\ref{eq:electric_field}.
The scaling is summarized in Table~\ref{tab2}.

\subsection{Modeling of pair cascades}
% Particle decay + thresholds
Here, we adopt the algorithm by \citet{Philippov2020,Cruz2021a,Cruz2022} based on a pair-production threshold energy $\gamma_\mathrm{thr} m_\mathrm{e} c^2$, where $\gamma_\mathrm{thr}$ is the threshold Lorentz factor, $m_e$ is the electron mass, and $c$ is the speed of light.
The quantum electrodynamic process is approximated by the production of a new electron--positron pair whenever the Lorentz factor of the primary particle exceeds $\gamma_\mathrm{thr}$.
A fraction of the kinetic energy of the primary particle is converted through a virtual $\gamma$-ray photon into the electron--positron rest mass and their kinetic energies in the same time step.
Both secondary particles are created having the same kinetic energy because that is the most probable outcome of the pair production process \citep{Erber1966}.
Their velocity vector has the same direction as the velocity vector of the primary particle, thereby fulfilling the momentum conservation requirements.
As demonstrated in the algorithm by \citet{Philippov2020,Cruz2021a}, the approach can be considered valid as long as the relation 
\begin{equation} \label{eq:energy_condition}
E_\mathrm{p}' \gg E_\mathrm{s}' \gg E_\mathrm{b}'
\end{equation}
holds for the kinetic energies of the primary particle $E_\mathrm{p}'$, secondary particle $E_\mathrm{s}'$, and background plasma particle $E_\mathrm{b}'$.

% Estimated acceleration length to gamma_threshold and comparison with real situation
We assumed a threshold Lorentz factor $\gamma_\mathrm{thr} = \zeta \gamma_\mathrm{thr}' = 2\times10^{7}$.
By scaling the threshold $\gamma_\mathrm{thr}'$ as $\sim \zeta^{-1}$, we keep the distance at which a charged particle is accelerated to the pair-creation threshold as invariant between both the real and the simulation scale polar cap (for uniform and constant electric field).
Because the electric field also scales as $E' = E / \zeta$, the particle is accelerated to $\gamma_\mathrm{thr}'$ in approximately the same time interval and distance when the particle's $\gamma \gg 1$, as the traveled distance is then $\Delta x \approx c \Delta t$, the particle's momentum changes as $\Delta p \approx q E\Delta t$, and the particle's $\gamma$ is $\approx p / mc$.
Hence, the ratio between the acceleration distance and the polar cap radius in our simulations is the similar to that of the real polar cap.
The amount of energy lost by the primary particle with $\gamma > \gamma_\mathrm{thr}$ is chosen to be a constant value $5\times10^{-4}$ of the primary particle kinetic energy in the unscaled (physical)  system. 
With that choice we ensure that the condition in Eq.~\ref{eq:energy_condition} is also fulfilled in the scaled (simulation) system.

% Radiative losses
The macro-particles undergo curvature radiative energy losses by a radiative reaction force in the form \citep{Jackson1998,Daugherty1982,Timokhin2010}
\begin{equation} \label{eq-RL}
    \left(\frac{\mathrm{d}p}{\mathrm{d}t}\right)_\mathrm{RR} = \frac{2 q^2}{3 m c} \frac{p^4}{\rho^2},
\end{equation}
where $p = \gamma \beta$ is the dimensionless particle momentum per unit of mass, $q$ is the particle change, $m$ is the particle mass, and $\rho$ is the curvature radius.
The numerical implementation follows the algorithm of radiative losses by \citet{Tamburini2010}.
We note that the particle mass and charge in Eq.~\ref{eq-RL} must be their realistic values as was shown by \citet{Vranic2016}.
We choose the curvature radius to be the same as the light cylinder radius, $\rho = c / \Omega = 1.2\times10^{9}$\,cm, and uniform for the whole polar cap.
\footnote{It is quite unlikely that the magnetic field in the polar cap would be a perfect dipole. 
Even relatively small deviations of the magnetic field from a perfect dipole would result in the breaking of the symmetry of the magnetic field and could prevent the radius of the curvature of the magnetic field lines from becoming infinite, as is the case around the dipole axis. 
Our model qualitatively mimics polar caps with a slight non-dipolar component in that we neglect the variations of the radius of curvature. 
We discuss the effects of varying radius of curvature in Sect.~\ref{discuss4.1}.}
For the calculation of the curvature energy loss in Eq.~\ref{eq-RL}, our implementation scales up the simulation particle momentum $p'$ to obtain the momentum of a real system $p$ as 
\begin{equation} \label{eq:momentum_scale}
p' \rightarrow \gamma' \rightarrow \gamma \rightarrow p,
\end{equation}
where $\gamma = \gamma' \zeta$.
The momentum $p$ is then used in the particle pusher to calculate the radiative energy loss via Eq.~\ref{eq-RL} in addition to the threshold depending pair production loss to yield a new momentum $p_\mathrm{new}$ that is then scaled back in the next step by the inverse procedure to Eq.~\ref{eq:momentum_scale} to provide the new simulation momentum $p_\mathrm{new}'$.
The initial momentum vector $\boldsymbol{p}'$ and the resulting momentum vector $\boldsymbol{p}_\mathrm{new}'$ are parallel.
We find that particles would be able to reach Lorentz factors about $\gamma \sim 10^{8}$ with this curvature radius when the pair-production loss described above is turned off.

%%%%%%%%%%%%%%%%%%%%%%%%%%%%%%%%%%%%%%%%%%%%%%%%%%%%%%%%%%%%%%%%%%%%%%%%%%%%%%%%%%%%%%%%%%%%%%%%%%%%
\section{Results}  \label{sec:results}

\begin{figure*}[ht!]
    % \centering
    \includegraphics[width=\textwidth]{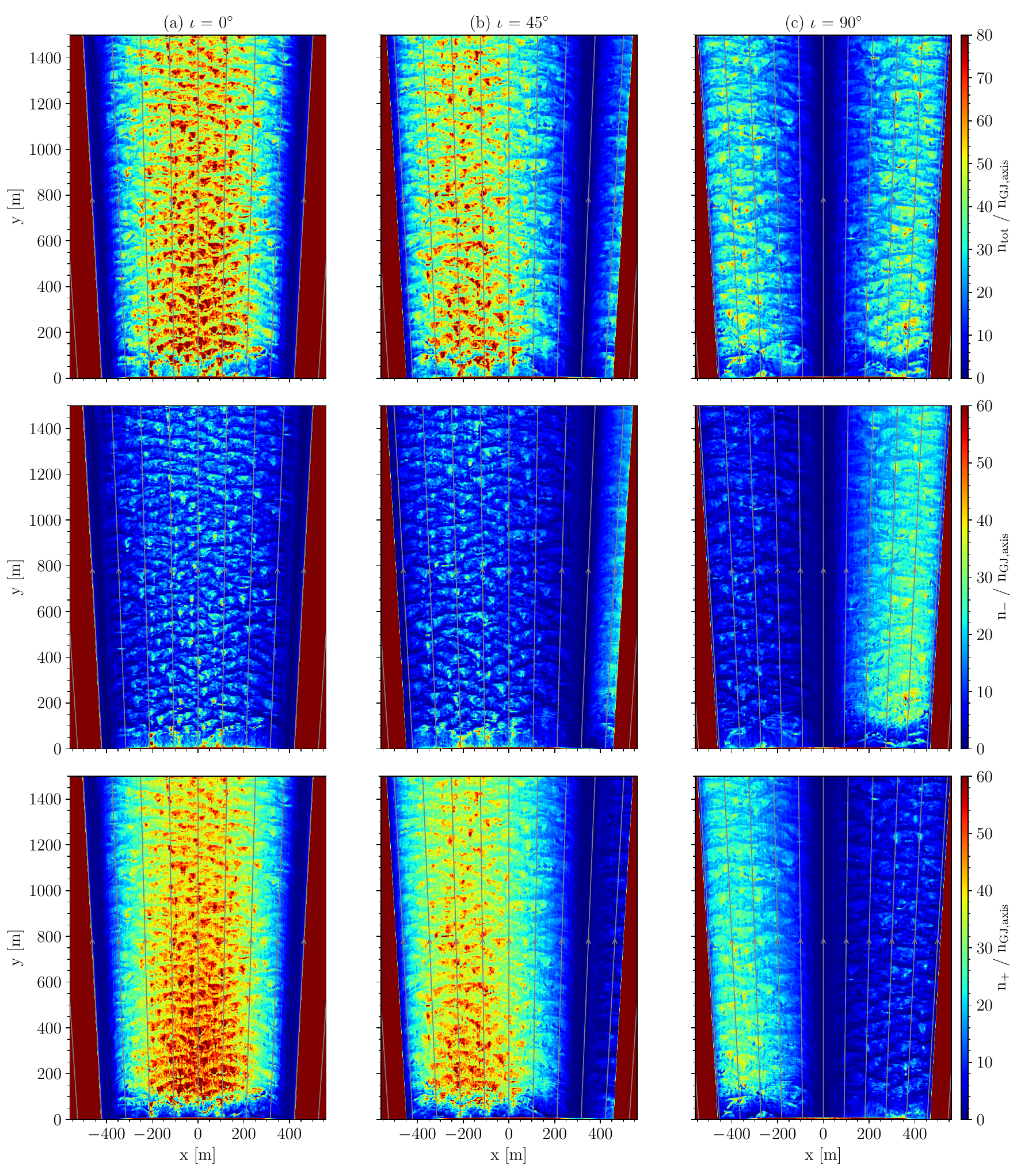}
    \caption{
            Total (top row), electron (middle row), and positron (bottom row) plasma density at the end of the simulation time normalized to the GJ density, $j_\mathrm{GJ,axis}$. 
        The inclination increases from left to right. 
        The magnetic field lines are denoted as gray lines.
        A video animation is on Zenodo.$^{2}$
        \label{fig2}
    }     
\end{figure*}

\begin{figure*}[ht!]
    % \plotone{fig3.pdf}
    \centering
    \includegraphics[width=\textwidth]{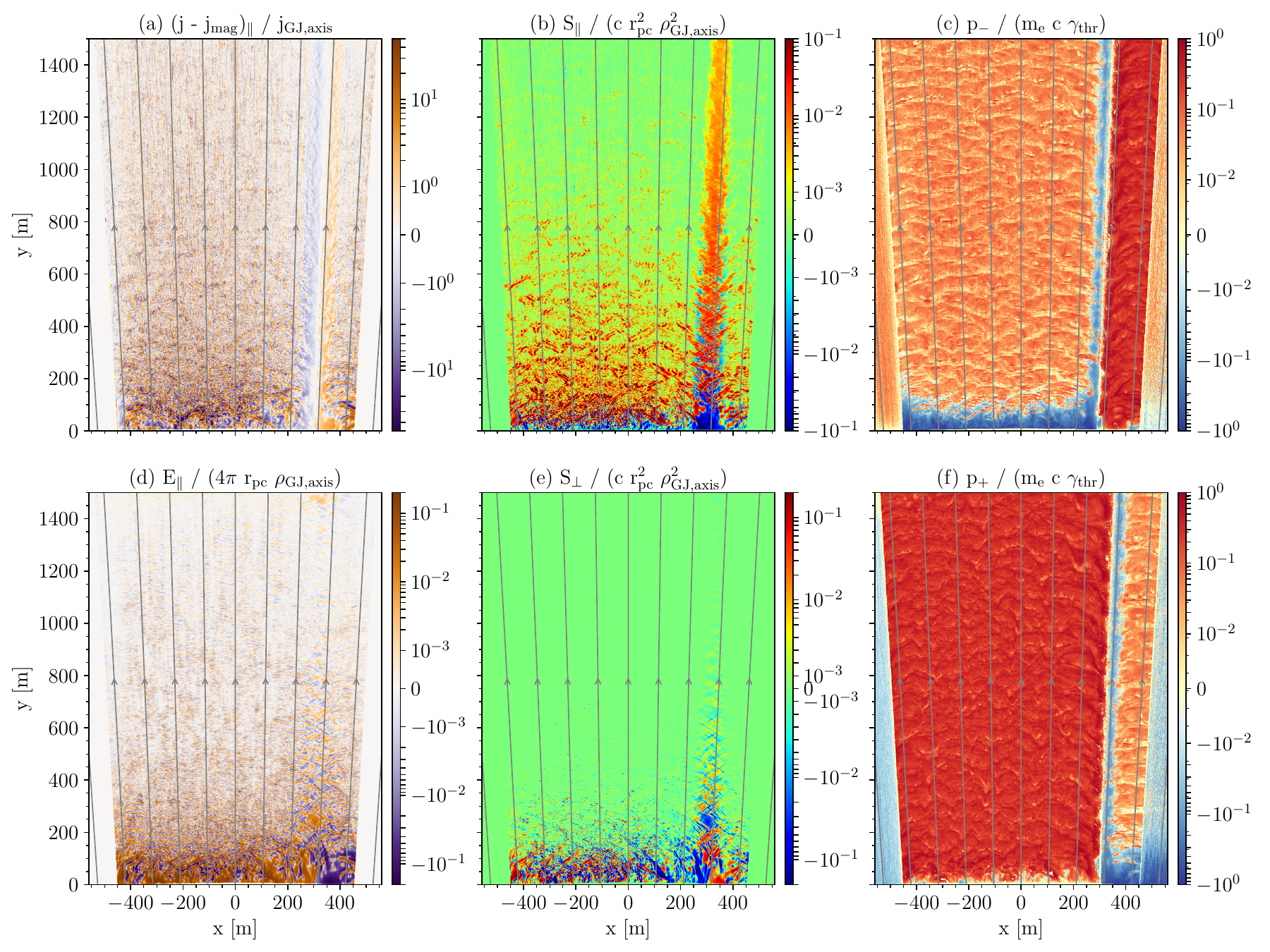}
    \caption{
        Plasma properties for the pulsar with $\iota = 45^{\circ}$. Panel (a): Parallel electric current densities to the external magnetic field. Panel (b): Parallel Poynting flux. Panel (c): Parallel electron bulk velocities. Panel (d): Parallel electric fields. Panel (e): Perpendicular Poynting flux. Panel (f): Parallel positron bulk velocities.
        The magnetospheric currents have been subtracted from the electric currents.
\label{fig10} 
}
\end{figure*}

\begin{figure*}[ht!]
    % \plotone{fig6.pdf}
    \includegraphics[width=\textwidth]{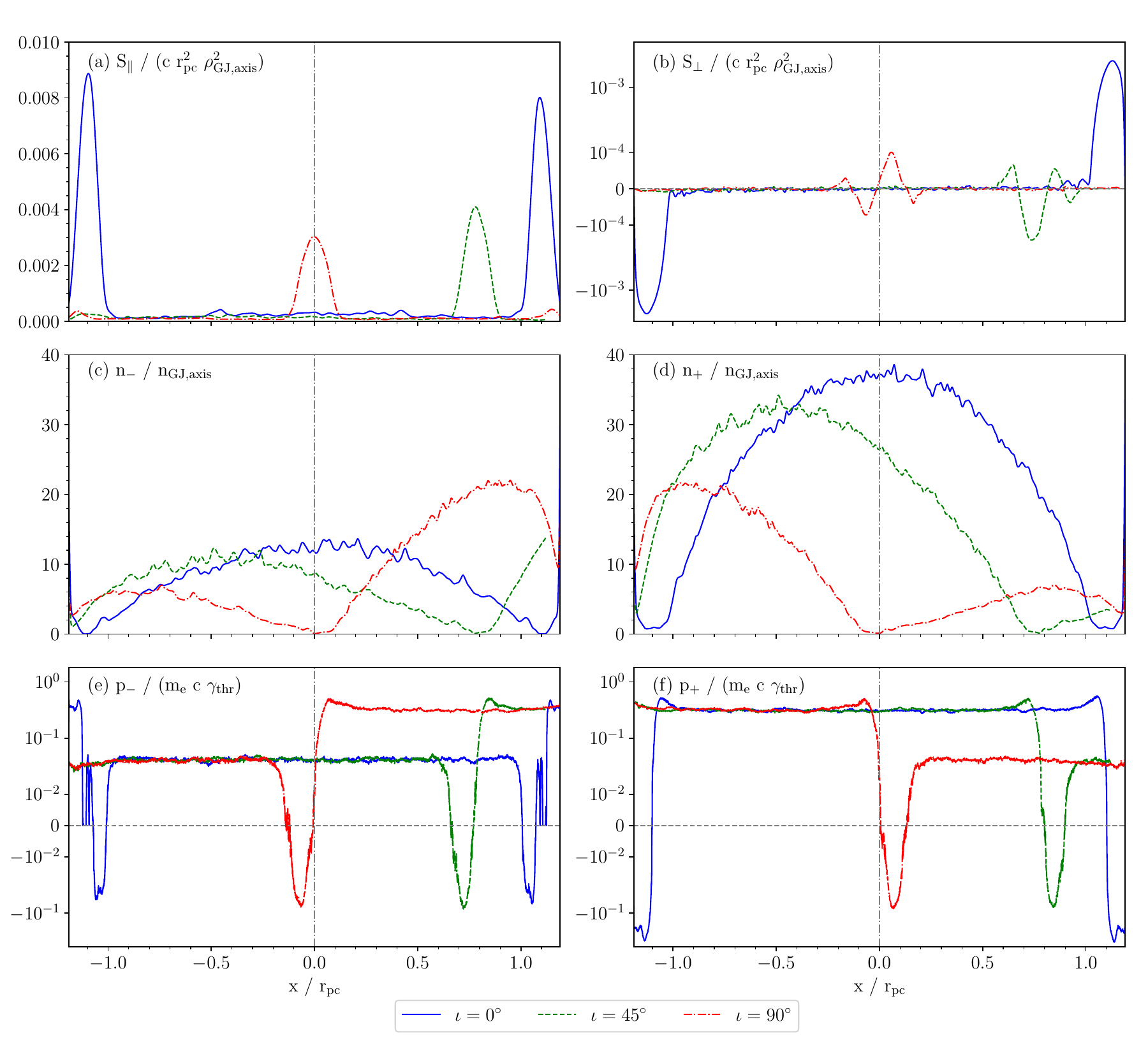}
    \caption{Time-averaged plasma quantities across the polar cap in the $x$-axis for $y = 11\,000\,\Delta_\mathrm{x} = 1474$\,m. 
    Time average between 130\,000 and 150\,000 time steps.
        Top row: Poynting flux parallel and perpendicular to the external magnetic field. 
    Middle row: Electron and positron plasma density. 
    Bottom row: Momenta of electrons and positrons. 
        }
    \label{fig6}
\end{figure*}

\begin{figure*}[ht!]
    % \plotone{fig8.pdf}
    \includegraphics[width=\textwidth]{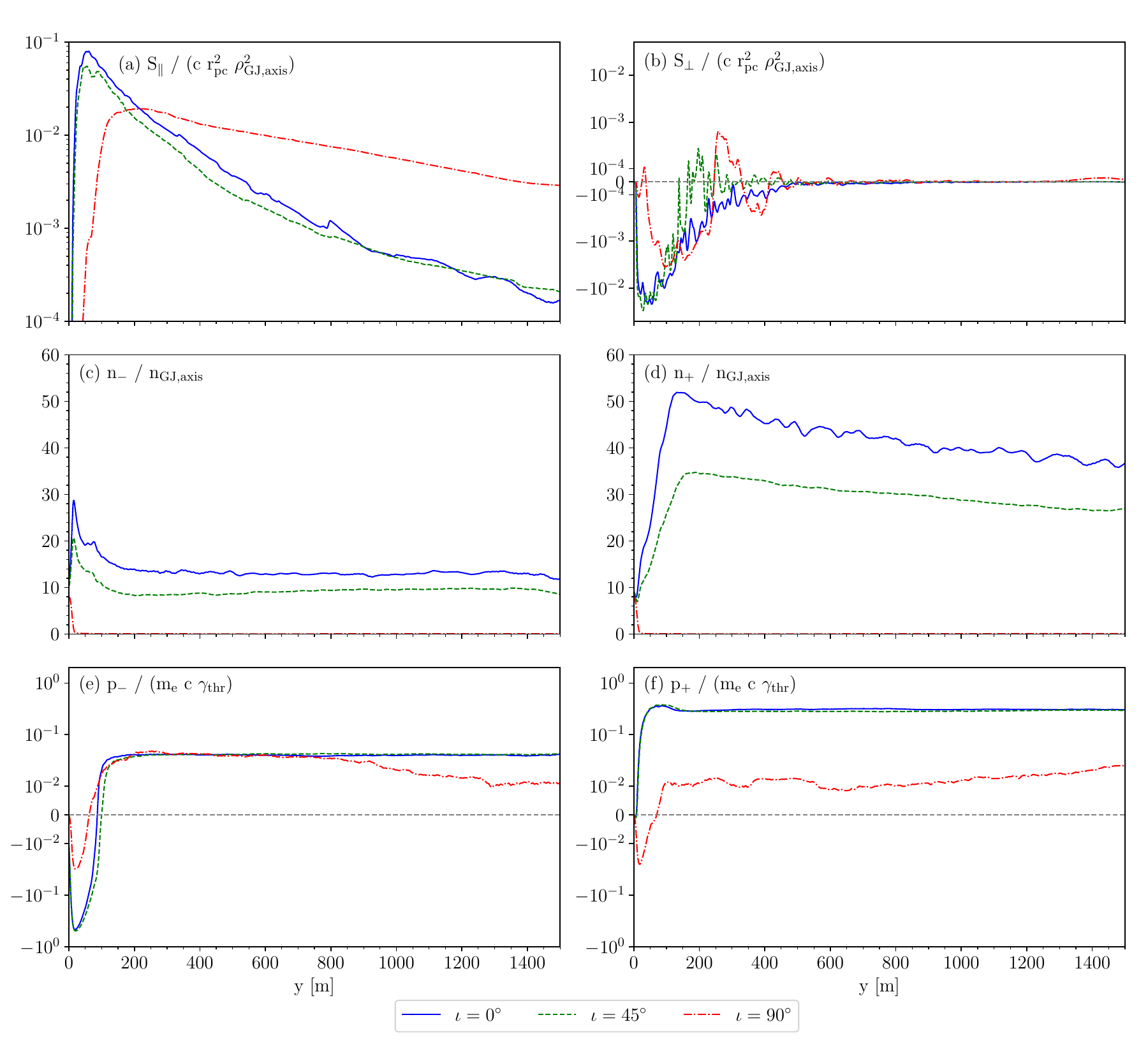}
    \caption{Same as Fig.~\ref{fig6} but for a cut along the dipole axis and $x = 0$.
    }
    \label{fig7}
\end{figure*}

\begin{figure*}[ht!]
    % \plotone{fig7.pdf}
    \includegraphics[width=\textwidth]{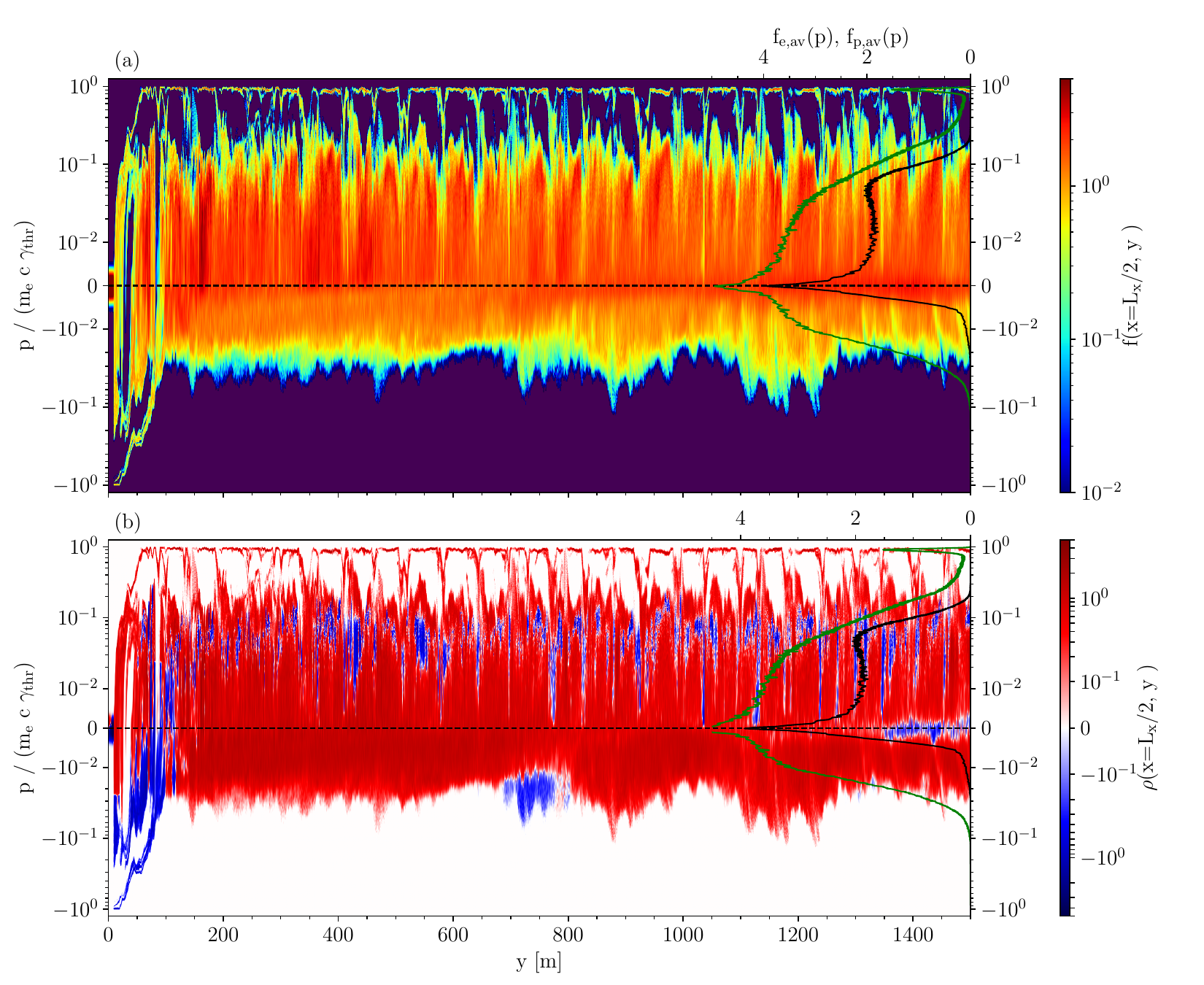}
    \caption{Particle phase space distributions for the pulsar with $\iota = 0^\circ$ along the magnetic dipole axis at $x=0$ at the end of the simulation time. 
        Black and green lines: Average particle distributions of electrons $f_\mathrm{e,av}(p)$ and positrons $f_\mathrm{p,av}(p)$, respectively, in a distance range of $y = 800$--1400\,m.
    Top: Particle number density. 
    Bottom: Charge density. The white regions have a zero charge density distribution (if $f(r,p) > 0$), or there are no particles (if $f(r,p) = 0$).
    }
    \label{fig8}
\end{figure*}

\begin{figure}[ht!]
    % \plotone{fig9.pdf}
    \includegraphics[width=0.48\textwidth]{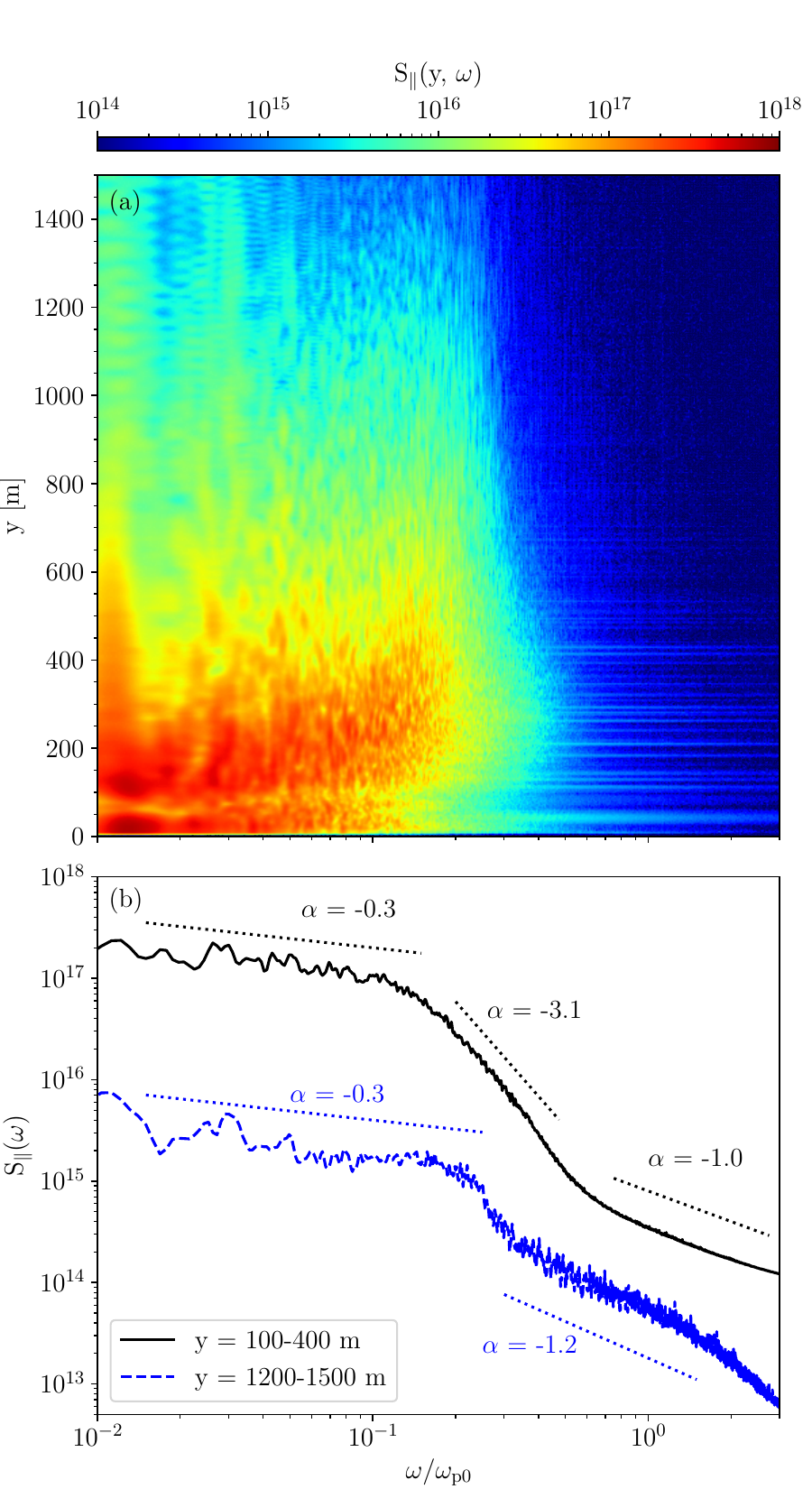}
    \caption{Spectrum of the parallel Poynting flux for the pulsar with $\iota=90^\circ$ at the dipole axis between time steps 130\,000--150\,000. Top: Spectrum as a function of the pulsar distance. Bottom: Average spectrum for two ranges of distances from the star surface. Power-law (dotted) lines and indices are denoted.
    }
    \label{fig9}
\end{figure}

After the simulation starts, the polar cap goes through two distinct evolutionary phases before it settles into a quasi-periodic discharge formation in the gap close to the star surface.
The described simulation setup is designed to go through these stages effectively and relatively fast.
The notation of scaled quantities by primes is omitted in this section --- all quantities associated with the skin depth, plasma density, and electromagnetic fields are scaled.
The time and space quantities remain the same, that is, $t = t'$, $x = x'$, and $y=y'$.

\subsection{Initial convergence of polar cap evolution}
In the first phase, the parallel electric fields grow because of the magnetospheric currents described by Eq.~\ref{eq:electric_field}.
For a video, please see \href{https://zenodo.org}{Zenodo}\footnote{\href{https://doi.org/10.5281/zenodo.11402572}{https://doi.org/10.5281/zenodo.11402572}}.
After the simulation start, the plasma particles are accelerated by the electric field until they reach the threshold Lorentz factor and begin to produce pairs.
It takes several hundred time steps for the first particles to reach the threshold Lorentz factor.
New pairs are created until the plasma currents sustain the magnetospheric currents in Eq.~\ref{eq:electric_field}, and the electric intensities decrease far enough so that particles cease being accelerated.
The polar cap regions where the magnetospheric currents are close to zero are not, or only slightly, populated by secondary particles.
When the pair production decreases and approaches zero at the end of this phase, the plasma density is typically much larger than the GJ plasma density and
the plasma particles are relatively smoothly distributed in space.

Perpendicular electric fields are mainly generated when secondary particles are being produced.
The secondary particles are generated in pairs, leading to an initial local charge neutrality; however, the pairs will be quickly separated when strong electric fields are present.
The randomly distributed formation of the charged particles across the magnetic field leads to the formation of electric fields with perpendicular and parallel components to the magnetic field lines.
The perpendicular fields do not interact with the particles via wave--particle interactions because the particles cannot move across the field lines.
The waves associated with perpendicular electric fields propagate at velocities close to the speed of light and mostly along the magnetic fields. 
They are eventually absorbed on the star surface or leave the open magnetic field region.
At the end of the first phase, the polar cap is filled with plasma that can sustain the magnetospheric currents, the electric fields are close to zero, and the pair production has ceased.

In the second phase, the plasma flows out of the simulation domain until there is not enough plasma to carry the magnetospheric currents.
When the plasma cannot  sustain the full magnetospheric currents anymore, the electric fields begin to grow again, and new secondary particles are generated in two transient gap regions of open magnetic field lines close to the simulation boundaries ---
close to the star surface and at the upper boundary that is opposite to the surface.

The plasma density is small in these transient gaps, and the parallel electric fields grow fast.
The pair production restarts, and quasi-periodic cascades of plasma bunch generation establish themselves.
The resulting bunches have relatively regular shapes in space, being caused by the smooth profile of magnetospheric currents and plasma density.
The bunches flow away from the gap regions and move inward from the boundaries of the simulation domain, their shapes changing as they propagate through the simulation domain.
The discharges are, however, suppressed when the first generation bunches reach the opposite gaps in the simulation.

The perpendicular electric fields have higher amplitudes in the second stage than in the first because the plasma particles are organized into larger density structures associated with local non-zero charge densities.
The perpendicular electric fields reach the same intensity as the parallel electric fields at the bunch edges, but the bunch production at the upper boundary opposite to the surface decreases.
A Poynting flux associated with oscillating electric and magnetic field components, $\boldsymbol{S} = \frac{c}{4\pi} \boldsymbol{E} \times \delta \boldsymbol{B}$, is produced at the bunch edges.
Please note that the time-averaged electric field is zero. 
The only electric fields present are the perturbed ones, $\boldsymbol{E} = \delta \boldsymbol{E}$, where $\delta \boldsymbol{E}$ is the perturbation.
In contrast, the magnetic field time-average is the magnetic dipole $\boldsymbol{B_\mathrm{dip}}$ with a perturbation $\delta \boldsymbol{B} \neq \boldsymbol{B} \equiv \boldsymbol{B}_\mathrm{dip} + \delta \boldsymbol{B}$.

The system is finally quasi-stable after these two phases, and only an oscillating gap close to the star surface remains as a source of electron--positron pairs.
This quasi-periodic state of the polar cap is reached after approximately $\approx 110\,000$ time steps, $\sim 4.1L_\mathrm{y}/c $. 
We continue the simulation for an additional 40\,000 time steps, $1.5L_\mathrm{y}/c$, in order to reach the final quasi-equilibrium state.
Figures~\ref{fig2}--\ref{fig5} show snapshots of plasma parameters in the simulation domain at the end of the simulation.

\subsection{Plasma properties in quasi-periodic state}
% Figure 2 - Snapshot of electron + positron density
Figures~\ref{fig2}--\ref{fig5} present the final plasma properties in the polar cap for all three inclination angles at the end of the simulation.
The figures are overlaid by the dipole magnetic field lines.
Because weak numerical fluctuations keep forming close to the upper boundary $y = L_\mathrm{y}$, we omit that part of the simulation domain from the figures.

Figure~\ref{fig2} shows the total, electron, and positron plasma density normalized to the GJ number density at the dipole axis on the star surface for an aligned pulsar, $n_\mathrm{GJ,axis}$.
The positron density is larger than the electron density almost everywhere for all three dipole inclinations, in particular on the open field lines and in all regions of negative magnetospheric current where the ratio $j_\mathrm{mag}/j_\mathrm{GJ,axis}$ is positive.
The electron densities increase, and the corresponding positron densities decrease only in the gap close to the star surface $y = 30-150$\,m; 
please, see the electric fields and Poynting fluxes below for more details about the gap localization.
Because the magnetospheric current switches polarity across the polar cap (Fig.~\ref{fig1}), the density ratio of electrons and positrons also changes.
Those open field lines where the magnetospheric current density goes to zero are filled with significantly fewer particles, thereby forming density cavities.
The plasma density in the closed field lines turns out to be much larger than the GJ plasma density.

In the return current region of open magnetic field lines close to the star surface, the first few cascades develop toward the surface in agreement with \citep{Timokhin2010,Timokhin2013}. 
However, after a few discharge cycles, the direction of the cascades changes to the opposite one, with discharges moving outward. 
This is probably caused by the particle injection scheme used in our simulations when we inject both positrons and electrons at the stellar surface. 
Electromagnetic waves in discharges are generated in both directions with roughly equal fluxes \cite{Philippov2020}. 
As our study concerns mostly with the distribution of the Poynting flux of these waves, the direction of the cascade propagation in the return current regions should not affect the overall results.

% Figure 3 - Parallel currents parallel + electric field, 
As an example of plasma properties, we consider the pulsar with $\iota = 45^{\circ}$ in Fig.~\ref{fig10}.
More detailed plasma properties as well as the used normalizations are described in Appendices~\ref{app:A}--\ref{app:C}.
The figure shows the electric current density, parallel electric fields, parallel and perpendicular Poynting flux, and parallel electron and positron bulk velocities.
We highlight here only selected properties.
The parallel electric fields and perpendicular Poynting flux are detected mainly in the gap region close to the star surface, being consistent with the above estimated size of the gap.
The parallel electric fields and parallel Poynting flux form a channel located in the low-density region, $x \approx $300--400\,m, where not pair production occurs as shown in Fig.~\ref{fig2}.
Along the channel, intense parallel Poynting flux is transported from the gap away.
In open field lines where higher plasma density than in the channel, the parallel Poynting flux decreases with the distance from the star.
Moreover, the positrons have larger bulk momenta than electrons in the regions $x \lesssim 300$\,m of positive current ratio $j_\mathrm{mag}/j_\mathrm{GJ,axis}$, and vice versa for the opposite sign of the magnetospheric current.

\subsection{Time-averaged profiles across and along the polar cap}
Figures~\ref{fig6} and \ref{fig7} show time-averaged profiles of the Poynting flux, plasma density, and particle momenta across the open magnetic field lines and along the dipole axis ($x = 0$), respectively.
The quantities are normalized the same way as in Figs.~\ref{fig2}, \ref{fig4}, and \ref{fig5}.
The profiles are obtained as time-averaged profiles from the last 20\,000 times steps.
This time interval is estimated to be long enough to cover the generation of $\approx 6-12$ consecutive bunches.
The data were retrieved from line cuts across the polar cap and along the dipole axis for every 20-th time step with a width of one grid cell.

The profiles across the simulation domain in $y = 11\,000\,\Delta_\mathrm{x} = 1474$\,m and along the $x$-axis are presented in Fig.~\ref{fig6}.
The Poynting fluxes are enhanced in regions with low magnetospheric currents (Fig.~\ref{fig1}).
That is, the highest fluxes shift from the polar cap edges for $\iota=0^\circ$ to the dipole axes for $\iota = 90^\circ$.
The ratio between electron and positron densities depends on the magnetospheric current.
The densities have a maximum inside the polar cap and a decrease toward the last open magnetic field lines.
For $\iota = 90^{\circ}$, the profiles are symmetrical between left--right side of the figures ($x < 0$ and $x > 0$, respectively) for parallel Poynting flux and plasma densities.
Also, the profile is antisymmetrical for perpendicular Poynting flux.
The profiles in the closed field lines are not drawn in Fig.~\ref{fig6}.

In Fig.~\ref{fig7}, the average Poynting flux amplitude increases with the distance close to the star, but beyond the gap, the flux decreases.
The average plasma density is very close to zero because there is only a limited number of particles along the dipole axis for $\iota = 90^\circ$.
For other inclinations, the electron density decreases with the distance from the star surface up to a distance of $\sim200$\,m, remaining approximately constant at larger distances.
The positron density increases in the gap region with the distance up to $\sim200$\,m and 
further away, it decreases with distance.
The bulk velocities of both species increase with the distance in the gap region but remain approximately constant further out.

\subsection{Particle phase space and wave spectrum along the dipole axis} 
\label{sect:3.4}
% Figure 7 - Phase space along the dipole axis
Figure~\ref{fig8} shows the phase space distribution of the particle number and charge densities along the dipole axis for $\iota=0^\circ$
in a selected region of a width of $4d_\mathrm{e}$.
The plots are overlaid by electron (black line) and positron (green line) distribution functions $f_\mathrm{e,av}(p)$ and $f_\mathrm{p,av}(p)$ averaged in a distance range $y = 800$--1400\,m.
The distributions are normalized as
\[
   \int_{-1}^1 \frac{(f_\mathrm{e,av} + f_\mathrm{p,av})}{m_\mathrm{e} c \gamma_\mathrm{thr}} d\mathrm{p} = 1.
\] 

Primary particles, their Lorentz factors being close to the threshold decay Lorentz factor, are evident in the upper parts of both subfigures, where the momenta are $p / (m_\mathrm{e} c \gamma_\mathrm{thr}) \sim 1$.
The particles reach the $p / (m_\mathrm{e} c \gamma_\mathrm{thr}) = 1$ at distances of 100--200\,m from the star in the gap region.
The phase space distributions show regular phase space holes around $p / (m_\mathrm{e} c \gamma_\mathrm{thr}) \approx 0.2 - 0.8$ and distances $>50$\,m.
The holes are created for both the primary and secondary particle distributions and are associated with the discharge modulation of the primary particles occurring in the polar cap.
The plasma has a relativistically broad distribution ranging from negative values of $p / (m_\mathrm{e} c \gamma_\mathrm{thr}) \approx -0.05$ to positive once $p / (m_\mathrm{e} c \gamma_\mathrm{thr}) \approx 0.2$ in the broad band where the momenta are below the maximum for the holes.
Most of the charge density phase space is dominated by positive charges; however, there are regions of negative density inside the plasma bunches.
Some particles flow back on the star at $y = 0-100$\,m and $p / (m_\mathrm{e} c \gamma_\mathrm{thr}) \approx -1$ to $-0.1$, being composed mostly of electrons (the blue structures).

The seed particles for the cascades can be the electrons and positrons from the surface but also from the induced negative electron flow toward the star surface (see Fig.~\ref{fig8}(a) for $y=0$\,m to $100$\,m, $p/m_ec\gamma=-1$ to -0.1).
This particle flow may also reach the threshold Lorentz factor for negative momenta and can produce secondary pairs close to the star surface.
Even though such a new pair may initially move toward the star surface, one of the pair particles will be accelerated away from the star in the strong gap field, thus creating the cascade in the outward direction.
The particle with the opposite charge quickly reaches the star surface, where it is absorbed. 
Because the deceleration distance of the secondary particles moving initially toward the surface is much smaller than the acceleration distance that is required for the production of curvature photons, primary particles with the right sign injected from the surface may participate in the pair creation mechanism just as well as all those secondary particles initially moving toward the surface and being reflected by the electric field.
We tested this process in a simulation without surface injection, and the cascades are indeed started just from the charges moving toward the surface.

% Figure 9 - Wave analysis of the large Poynting flux for 90 deg
Figure~\ref{fig9} shows the spectrum of the outgoing Poynting flux at the dipole axis of one grid cell width for the Poynting flux channel for the pulsar with inclination $\iota=90^\circ$.
The spectrum is averaged over time steps 130\,000--150\,000 ($\approx$$26.1-30.2\,\mu$s).
Only positive values were chosen for the calculation of the outgoing parallel Poynting flux --- negative values of the Poynting flux occur approximately in the gap region.
The frequency is normalized to the initial, non-relativistically calculated plasma frequency, $\omega_\mathrm{p0}$.
Most of the spectral power is emitted below $\omega_\mathrm{p0}$ mainly because the plasma is relativistically hot and the local plasma frequency decreases as $\sim \langle 1 / \gamma^3 \rangle$.
All waves have higher intensities up to $y < 400$\,m than at larger distances where they are partially absorbed.
Most energy is absorbed in the region $\lesssim 0.2 \omega_\mathrm{p0}$.
For the distance $y = 100$--400\,m, the spectrum has three power-law parts with indices $-0.3$, $-3.1$, and $-1.0$, while for distance $y = 1200$--1400\,m, the spectrum has indices $-0.3$ and $-1.2$ with a small but steeper transition region between them at $\omega/\omega_\mathrm{p} \approx 0.2-0.3$.
We also found that most waves follow the vacuum dispersion relation of the electromagnetic waves $\omega \approx c k$ for almost all frequencies.

%%%%%%%%%%%%%%%%%%%%%%%%%%%%%%%%%%%%%%%%%%%%%%%%%%%%%%%%%%%%%%%%%%%%%%%%%%%%%%%%%%%%%%%%
\section{Discussion} \label{sec:discuss}

The polar caps of neutron star magnetospheres produce pair cascades and are sources of coherent radio emissions, ultrarelativistic particles, and plasma outflows.
We carried out 2D kinetic particle-in-cell simulations of the polar cap close to the star surface where the polar gap is located.
We studied how the inclination angle of the magnetic dipole influences the discharge behavior, bunch generation, and Poynting flux transport.

% Main results about Poynting flux transport in channels
The main driving mechanisms of the polar cap pair cascades are the magnetospheric currents and their polarity switch across the polar cap.
We found that the Poynting flux is effectively transported away from the gap and through the magnetosphere in regions of open magnetic field lines where the magnetospheric currents are close to zero, denoted as the Poynting flux channels.
The plasma only has a low density in these channels because of the weak or lacking secondary plasma generation.
The Poynting flux channel locations depend on the inclination angle of the dipole field.
For an aligned pulsar, there are two channels at polar cap edges close to the closed field lines.
The channels with larger latitudes diminish for increasing inclination angle, and the second channel shifts toward the dipole axis and reaches it for an orthogonal pulsar.

The width of the transport channel across the magnetic field lines is determined by the boundaries of the regions where the plasma can sustain the magnetospheric currents without significant pair production.
In the simulation, we imposed the condition for a minimal plasma density at the boundary that corresponds to the plasma frequency $\omega_\mathrm{p0}$, in our simulations represented by 2\,PPC of the boundary condition.
However, a choice of another minimal density, for example, by the GJ current density \citep{Beloborodov2008}, could change the channel width.

\subsection{Poynting flux properties} \label{discuss4.1}
%The Poynting flux is weakly absorbed in the channel because of the low plasma density in the region.
Because the plasma density is low in the Poynting flux channel, the electromagnetic wave frequencies are above the relativistically corrected plasma frequency, namely, $\sim (10^{-4}-10^{-3}) \omega_\mathrm{p0} \ll \omega_\mathrm{p0}$ (Fig.~\ref{fig9}), and the absorption in the channel is low.
The Poynting flux enters the channel in the gap region where it is very high, propagating from there out in various directions.
The parallel component of the flux dominates the perpendicular component by a few orders of magnitude after a certain propagation distance along the channel.
The perpendicular component is quickly absorbed by its reflections on the channel boundaries that have higher plasma densities than the channel.
Specifically, a part of the electric field component parallel to the external magnetic field can be absorbed because the plasma can partially screen out that electric field.
We also find that the electric and magnetic waves of the Poynting flux follow the electromagnetic wave dispersion relation in a vacuum $\omega = ck$, and the Poynting flux can therefore be interpreted as being composed of electromagnetic waves.

The spectra of escaping waves can be described by broken power laws, having a relatively flat low-frequency region and a steeper high frequency part, with respective indices of $-0.3$ and $-1.2$.
It is conceivable that the Poynting flux in the channels and the flux associated with the bunches in high-density regions can propagate out and leave the magnetosphere as electromagnetic waves that would then be observed as pulsar radio emission.

% Flux escape from the channel + collimation
Assuming that the Poynting flux channel also exists at a larger distance from the star than the length of the simulation domain, the Poynting flux can be captured in the channel until it reaches a distance at which the wave frequencies exceed the plasma frequency outside the channel.
Thus, the higher frequency waves could escape the channel earlier than the low-frequency ones because their frequencies earlier exceed the plasma frequency of the surrounding plasma.
For an observation at an arbitrary radio frequency, the channel could serve as a ``waveguide'' through the magnetosphere into the wave origin --- the gap.
By observing a low-frequency cutoff in radio frequencies, we could estimate the maximal allowed plasma density in the emission region --- that is, the plasma frequency and maximal density in the channel. 

% Similar to optical fiber
As the Poynting flux propagates through the channel, it may undergo reflections at the channel boundaries because the refractive index of the plasma increases toward the denser regions.
In addition, only the reflection of the Poynting flux component polarized perpendicular to the channel boundary is expected, similar to the reflection of a light wave in an optical fiber.
Thus, we hypothesize that if the Poynting flux escapes the channel as electromagnetic waves, its polarization can follow the channel's geometric structure.
This way, the observed polarization profiles of pulsars could be directly related to the geometrical structure of the Poynting flux channel in the polar cap.

We assumed a constant curvature radius across the polar cap. 
Hence, the pair production also occurs throughout the whole region where the current density allows efficient particle acceleration, for example, at the center of the polar cap for an aligned rotator. 
In the case of a quasi-dipolar magnetic field, when the axis of symmetry of the magnetic field is inside the polar cap, the radius of curvature of the magnetic field lines around the symmetry axis goes to infinity (for dipolar field $\rho \propto 1/\theta'$, $\theta'$ being the colatitude measured from the magnetic axis). 
This will lead to the suppression of pair formation in the region around the symmetry axis. 
The presence of a low-density plasma around the axis can therefore give rise to the formation of an additional Poynting flux channel. 
However, the size of this region should be relatively small, as the height $h$ of the accelerating gap depends on the radius of curvature quite weakly, $h \propto \rho^{2/7} \propto \theta^{-2/7}$ \citep[Eq.~40]{Timokhin2015}; so $h$ becomes high (indicating that the pair formation is suppressed) only for very small values of $\theta$.

Low-density magnetic channels, called flux tubes, similar to those in our simulations have also been found in theoretical papers from solar physics \citep{Wu2002,Schlickeiser2010,Treumann2014} and there have also been recent in situ detections \citep{Chen2023}.
An electromagnetic wave can propagate inside the flux channel until the surrounding plasma density decreases with the distance from the Sun and the wave frequency exceeds the plasma frequency of the surrounding material.

% Poynting flux absorption
In the regions of high magnetospheric currents, irregular pair plasma bunches and clouds of particles are formed in the gap close to the star surface of a thickness of $\sim150$\,m.
Though the plasma density significantly varies in the bunch, the bunch shape only weakly changes during its propagation along the magnetic field lines.
In our simulations, the Poynting flux associated with the bunches is absorbed in these regions by the plasma, in contrast with the low-density channels, and the flux decreases with the distance from the star.
We tested the Poynting flux decrease for various simulation setups and found the decrease to be independent of the simulation grid cell size and PPC as long as these parameters allow sufficient resolution of the polar cap and the released bunches.
However, the fact that we considered a scaled-down problem, where we do not resolve the actual plasma scales as well as detailed dynamics of pair formation, did not allow us to exclude the possibility that the electromagnetic emission can escape the dense plasma.  
Details of the wave propagation are unclear, especially when taking into account highly non-uniform and non-stationary plasma density distribution.
The waves associated with bunches might escape the magnetosphere as electromagnetic O-mode waves as suggested, for example, by \citet{Philippov2020} and \citet{Tolman2022}.
However, an important conclusion from our simulations is that even if the waves generated during discharges are absorbed in dense plasma regions, the low plasma density regions, where pair formation is suppressed, could serve as propagation channels for these waves, allowing them to escape the magnetosphere.

\subsection{Plasma bunch and gap properties}
% Bunch properties
In general, however, it is not straightforward to determine the typical distance between the bunches that are released from the gap region from our simulations.
While the gap in the simulations spreads out up to the distance of $\sim150$\,m from the star surface, similar to Sturrock's model \citep{Sturrock1971}, the consecutive bunches are closer together, being separated only by $\sim$50--100\,m.
The bunches are, moreover, not equidistantly distributed along or across the magnetic field lines.
Their distance depends on the strength of the magnetospheric current: typically, the gap size is larger for lower magnetospheric currents and therefore larger for lower dipole magnetic fields.
Hence, the distance varies with the magnetospheric current profile across the polar cap.

The bunch sizes and distances are also significantly influenced by the values of primary particle Lorentz factors at which they begin to emit $\gamma$-ray photons energetic enough to produce pairs.
This relation is, nonetheless, not a simple function of the plasma parameters. 
A lower number of secondary particles produced from lower $\gamma-$ray emission can lead to a lower secondary plasma density and lower plasma currents and can result in higher electric field intensities. 
The high-intensity fields may change the typical acceleration distance, gap size, and bunch distances.

The electric gap size may change when one includes the mean free path of curvature photons before they create secondary pairs.
The increase can occur because the photon mean free path is inversely proportional to the energy of curvature photons \citep{Timokhin2013,Cruz2021b}.
When the inclusion of the photon propagation leads to an increase of the gap size, the primary particles can be accelerated to larger energies in a larger gap (assuming that they do not reach their radiation reaction limited factor), leading to photons with higher critical energies and shorter mean free paths. 
However, if the primary particles reach their radiation reaction limited Lorentz factor when they produce the curvature photons, then they cannot produce photons with shorter mean free paths, and the gap size is larger than in our case.
Studying this effect requires the computation of the probability for curvature photon emission and the photon mean free path for each time step in the simulation.
We omitted this effect as the additional computational load would have made the problem intractable for our computational resources. 
Our results are therefore meant to qualitatively illustrate the local dynamics of the surface pair creation, the flux channel formation and subsequent radio emission. 
Nevertheless, more quantitative modeling may be feasible in the future when increased computational resources may allow for it.

\subsection{Comparison with previous 2D PIC simulation}
There is only one 2D PIC simulation of the polar cap that resolves realistic sizes of the gap region and the bunches that we could use for comparison with our results \citep{Cruz2021b}.
The structure of bunches, their distances, and sizes obtained from our simulations are different from theirs because of differences in the magnetospheric parameters and simulation setup.
The pair cascades in our simulations are driven by local magnetospheric currents formed as the result of the curvature of magnetic field lines.
The current profile, which is used across and along the polar cap, is a solution of global force-free general-relativistic simulations, and the profile changes with the dipole inclination angle.
The simulations by \citet{Cruz2021b} drive the cascade by electric fields present at the star surface and are implemented as the boundary condition for an arbitrary profile across the polar cap.
In addition, it is unclear whether the simulations were carried out long enough so that a quasi-steady state can evolve with the cascade.

The distances between bunches are smaller in our simulations because they depend on the strength of magnetospheric currents.
We assumed that the magnetospheric currents correspond to a pulsar with $P = 0.25$\,s and a dipole magnetic field of $10^{12}$\,G at the surface.

Another important process that can significantly increase the bunch distances, as showed by \citet{Cruz2021b}, is the $\gamma$-ray photon propagation distance before they decay into pairs.
We have not included this effect in our simulations.
For a more precise consideration of the QED effects of the $\gamma$-ray photon emission and decay, we expect a smoother density profile of the bunches, as both the photon emission and pair producing distances will have a distribution of values.
However, the probability distributions of the QED effects are included in our approach in the form of an appropriate step function at the threshold Lorentz factor. 
The maximal Lorentz factor that particles can reach strongly depends on the radiative energy losses.
It is, nonetheless, uncertain how high the Lorentz factors the primary particles can reach when all types of radiative losses (like curvature, synchrotron, and linear acceleration emission losses) are included.

\section{Conclusions} \label{sec:conclude}
We have investigated the pair cascades in the polar cap of neutron star magnetospheres through PIC simulations at kinetic scales, aiming to find out how the magnetic dipole inclination impacts the production of plasma bunches and the generation and possible escape of electromagnetic radio waves in the form of Poynting flux.
We find that the high intensity Poynting flux generated in the electric gap close to the star surface can be transported in channels of low plasma density through the magnetosphere without significant absorption.
The transport channels are located along open magnetic field lines where the magnetospheric currents approach zero \citep{Gralla2017}.
Oppositely, outside these channels, for example, at the edges of plasma bunch discharges, the generated Poynting flux decreases quickly with the distance from the star.

It is interesting to note that no conversion process from the kinetic energy of plasma particles to electromagnetic waves is necessary in order to emit a highly intensive radiation field, as the obtained broad spectrum of escaping electromagnetic waves is directly generated by the oscillating electric field gap.
This result is fundamentally different from that of most other theoretical mechanisms of pulsar radio emission, where  high conversion efficiencies of plasma waves to coherent radio emissions are required. \citep{Eilek2016,Melrose2020a,Philippov2022}.
The escaping electromagnetic waves have two power-law regions with indices -0.3 for $\omega/\omega_\mathrm{p} \approx 0.01-0.2$ and -1.2 for $\omega/\omega_\mathrm{p} \approx 0.3 - 2$.
For higher frequencies, the spectral index is similar to those observed in pulsars (for example, the mean value $\sim $-1.4 by \citet{Bates2013} and \citet{Bilous2016}) or estimated from analytical and numerical cascade modeling, such as $-1$ by \citet{Tolman2022}.

Our model implies that the pulsar radio beam does not have a symmetrical cone-like structure. Rather, the radio waves propagate from the gap with oscillating electric fields in channels along open magnetic field lines where no pair cascades occur and the plasma density is low.
Therefore, the shape of the pulsar radio beam is defined by the shape of magnetic field lines without pair production.
In our emission model, all radio frequencies are generated at the same time and in a confined region, and
we expect to receive them without any significant frequency-dependent delays apart from those caused by interstellar dispersion.
The observations of \citet{Hassall2012,Hassall2013} do not support the frequently conjectured frequency–to–radius relation in pulsar emission and may be seen as empirical support for our model.
The frequently conjectured frequency--to--radius relation of the escaping radiation has no basis in emission physics in our model.

The detected behavior of the Poynting flux poses the question of whether the Poynting flux outside or inside the Poynting flux channels can reach higher altitudes in the magnetosphere and leave the magnetospheric plasma as electromagnetic radio waves. 
We hypothesize that the flux in the form of electromagnetic waves can reach higher altitudes and is efficiently captured in the channel as long as the decreasing plasma frequency in the surrounding field lines is higher than the electromagnetic wave frequency. 
The waves could escape the channel beyond that point. 
In addition, if the observed radiation of pulsar pulses is transported in such a Poynting channel, the pulse profile and the pulse maxima will be correlated with the Poynting flux channel position and the field lines with very low or zero current.
In order to interpret the pulsars with more than two radio peaks, one has to assume more complex structures of the magnetospheric current across the polar cap than those given by a simple magnetic dipole \citep{Lockhart2019}.

Our numerical simulations have been found to be resilient to moderate changes in initial and boundary conditions;
these simulations can therefore allow for more detailed future studies of various aspects of the discharge evolution in the neutron star polar cap, such as star surface heating, high energy emissions, particle acceleration, various additional processes leading to radiative losses, and the quantum-electrodynamic pair creation.

%%%%%%%%%%%%%%%%%%%%%%%%%%%%%%%%%%%%%%%%%%%%%%%%%%%%%%%%%%%%%%%%%%%%%%%%%%%%%%%%%%%%%%%%%%%
\begin{acknowledgements}
We acknowledge the developers of the ACRONYM code (Verein zur F\"orderung kinetischer Plasmasimulationen e.V.). 
We also acknowledge the support by the German Science Foundation (DFG) projects MU~4255, BU 777-16-1, BU~777-17-1, and BE~7886/2-1. 
A.T. was supported by the grant 2019/35/B/ST9/03013 of the Polish National Science Centre.
The authors gratefully acknowledge the Gauss Centre for Supercomputing e.V. (\url{www.gauss-centre.eu}) for partially funding this project by providing computing time on the GCS Supercomputer SuperMUC-NG at Leibniz Supercomputing Centre (www.lrz.de), project pn73ne.
\end{acknowledgements}


\bibliographystyle{aa}
\bibliography{references}
%\bibliography{../../references}

\begin{appendix} 

{%\bf

% \section{Detailed plasma properties in quasi-period state} \label{app:A}
\section{Electric fields and currents} \label{app:A}

% Figure 3 - Parallel currents parallel + electric field, 
Electric current densities and electric fields parallel to the magnetic field lines are shown in Fig.~\ref{fig3} with
the current density being  normalized to the GJ current density, $j_\mathrm{GJ,axis}(\boldsymbol{B}_\mathrm{dip,axis})$.
The electric field is normalized by the radius of the polar cap, $r_\mathrm{pc}$, and the GJ charge density for an aligned pulsar at the dipole axis on star surface, $\rho_\mathrm{GJ,axis}(\boldsymbol{B}_\mathrm{dip,axis})$, in order to obtain a dimensionless quantity.
The strongest currents and electric fields are formed in the gap regions close to the star surface, with the electric field amplitudes decreasing with increasing distance from the star.
The currents and electric fields are, however,  not enhanced in regions of low magnetospheric current (Fig.~\ref{fig1}).
This happens as a consequence of too few particles being  available to sustain the magnetospheric currents, and also  insufficient numbers of particles that can be accelerated to produce enough pairs for quenching the electric fields.

The amplitude of the parallel electric field $E_{||}$ reaches values about $\approx6.6\times10^{10}\,\mathrm{V}\,\mathrm{m}^{-1}$ for all three inclination angles. 
Accelerating a particle at rest in the gap to the kinetic energy corresponding to $\gamma_\mathrm{thr}$ ($\approx$$10^{13}$\,eV) requires a distance of $150$\,m in this field, assuming that the electric field is constant in time and space during the acceleration.
This acceleration distance is consistent with the above estimated size of the gap.

% \vspace{-5cm}
\noindent
\begin{minipage}{1.0\textwidth}
  \strut\newline
  \centering
  \includegraphics[width=1.0\textwidth]{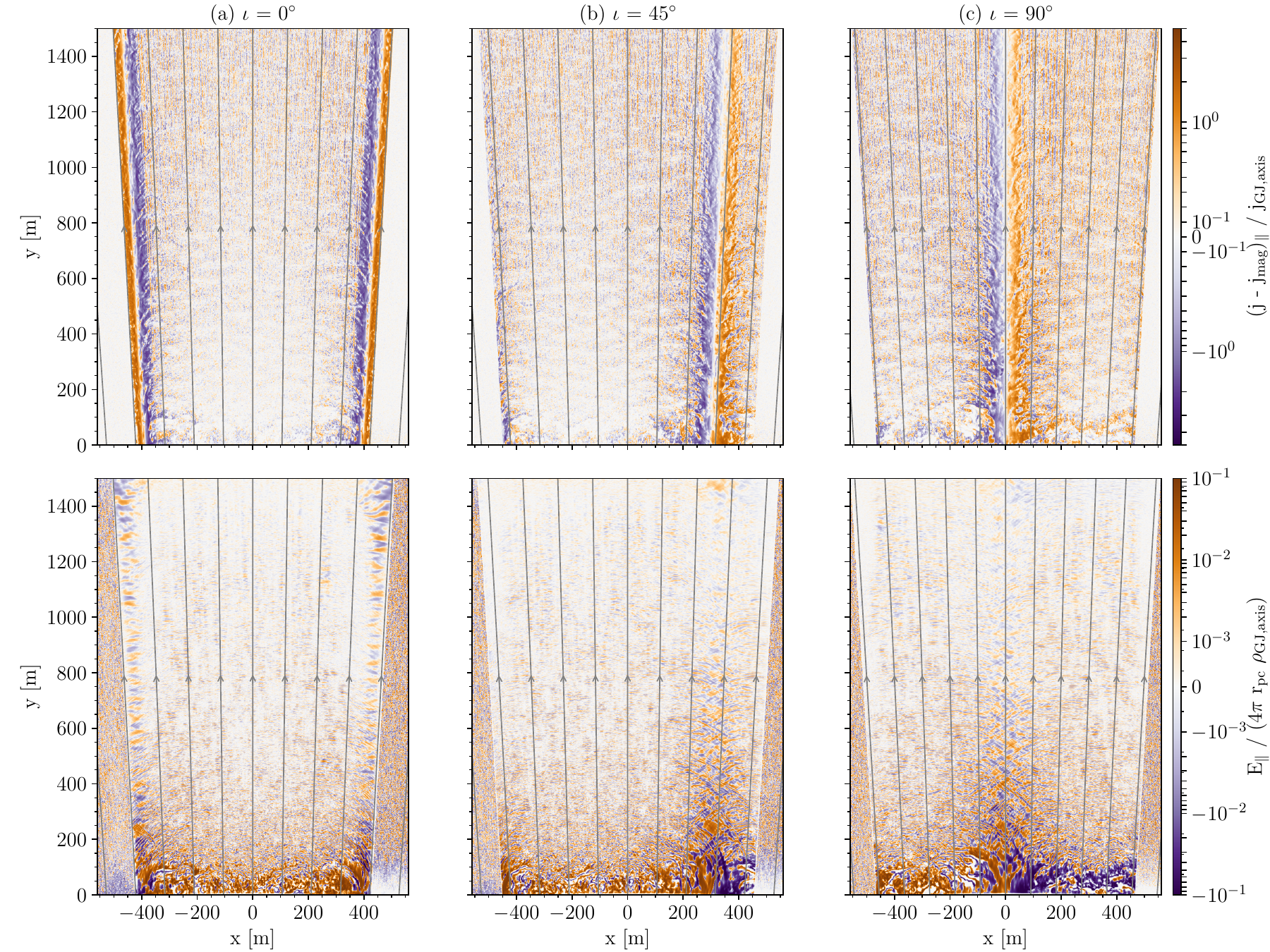}
  \captionof{figure}{Same as Fig.~\ref{fig2} but for parallel electric currents (top row) and parallel electric field (bottom row). The magnetospheric currents are subtracted from the electric currents.}\label{fig3}
\end{minipage}

% \FloatBarrier

\clearpage

\section{Parallel and perpendicular Poynting flux} \label{app:B}

%Figure 4 Poynting parallel + perp
Figure~\ref{fig4} shows Poynting flux parallel and perpendicular to the external magnetic field lines.
The Poynting flux is normalized to the radius of the polar cap, $r_\mathrm{pc}$, and the GJ charge density, $\rho_\mathrm{GJ,axis}(\boldsymbol{B}_\mathrm{dip,axis})$. 
The dominant electric field component of the parallel flux is perpendicular to the external magnetic field and lies on the $\boldsymbol{x} - \boldsymbol{y}$ plane.
For the perpendicular flux, the dominant component is the electric field parallel to the external magnetic field.
The dominant magnetic field components of the flux are associated with these electric fields, that is, perpendicular to both the corresponding electric field vector and the Poynting flux vector.
Hence, the Poynting flux is polarized in the simulation plane.

The parallel Poynting flux is larger than the perpendicular one as soon as one leaves the gap region.
The parallel flux is the largest in open field lines with low magnetospheric currents, where the plasma is not dense enough to absorb the flux.
Channels form in the regions of low magnetospheric currents, allowing the Poynting flux to be transported away from the star.
In contrast, the Poynting flux decreases significantly with the distance from the star in those open field lines with higher magnetospheric current.

\noindent
\begin{minipage}{1.0\textwidth}
  \strut\newline
  \centering
  \includegraphics[width=1.0\textwidth]{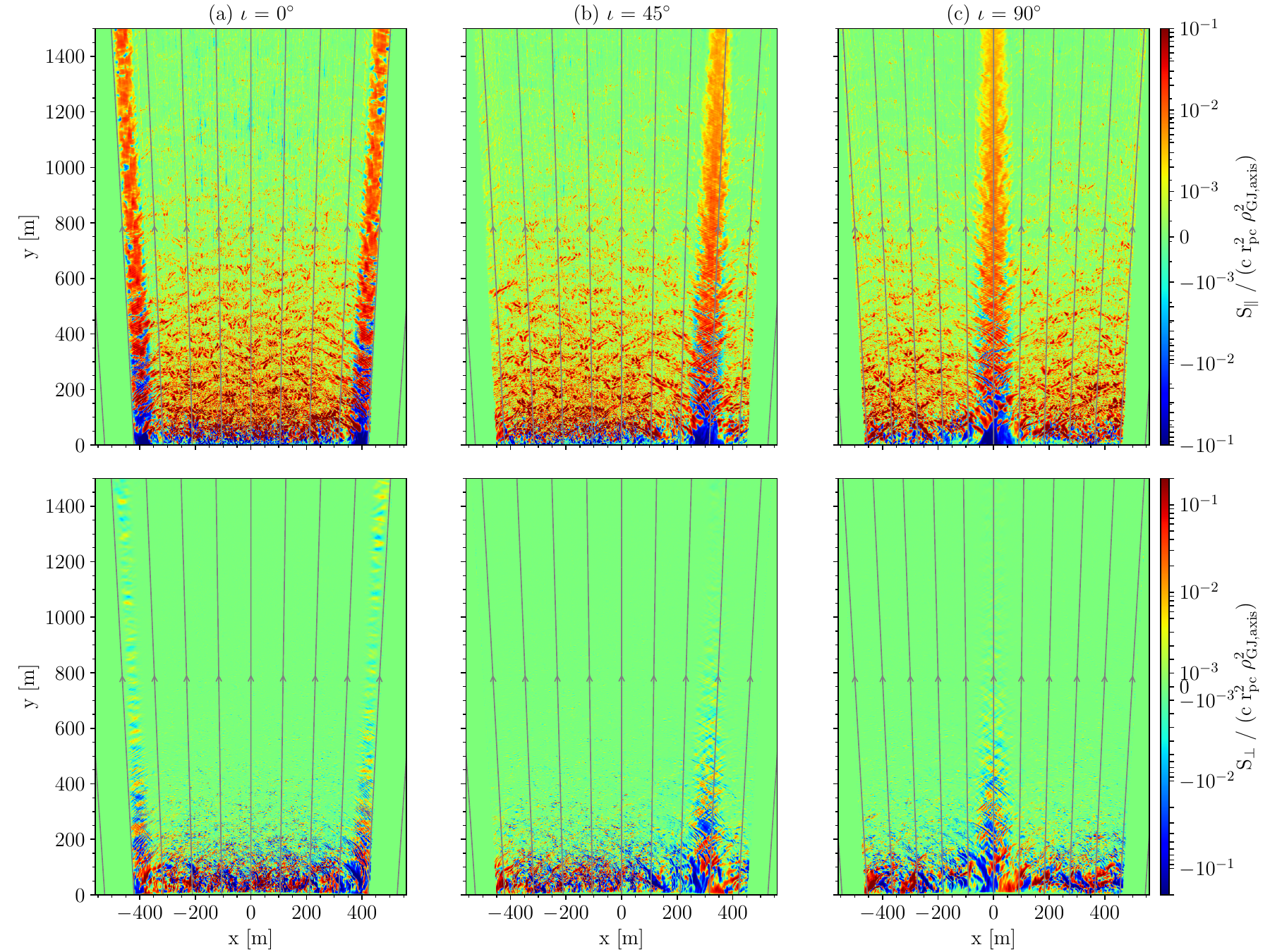}
  \captionof{figure}{Same as Fig.~\ref{fig2} but for the parallel (top row) and perpendicular (bottom row) Poynting flux to the external magnetic field at the end of the simulation time. 
  }\label{fig4}
\end{minipage}

\clearpage
% \FloatBarrier

\section{Electron and positron bulk velocities} \label{app:C}

% Figure 5 - bulk velocity of electrons and positrons
The bulk momenta of electrons and positrons are shown in Fig.~\ref{fig5}.
The bulk momenta are calculated as the mean momentum per particle of a given species averaged over the macro-particle weighting function.
The momenta are normalized to the electron mass, $m_\mathrm{e}$, light speed, $c$, and threshold Lorentz factor, $\gamma_\mathrm{thr}$.
The positrons have larger bulk momenta than electrons in the regions of positive current ratio $j_\mathrm{mag}/j_\mathrm{GJ,axis}$.
For the opposite sign of the magnetospheric current, the bulk momenta of electrons are larger than for positrons.
Both particle species have positive velocities at distances larger than $\sim$$r_\mathrm{gap}$$\sim$150\,m.

\noindent
\begin{minipage}{1.0\textwidth}
  \strut\newline
  \centering
  \includegraphics[width=1.0\textwidth]{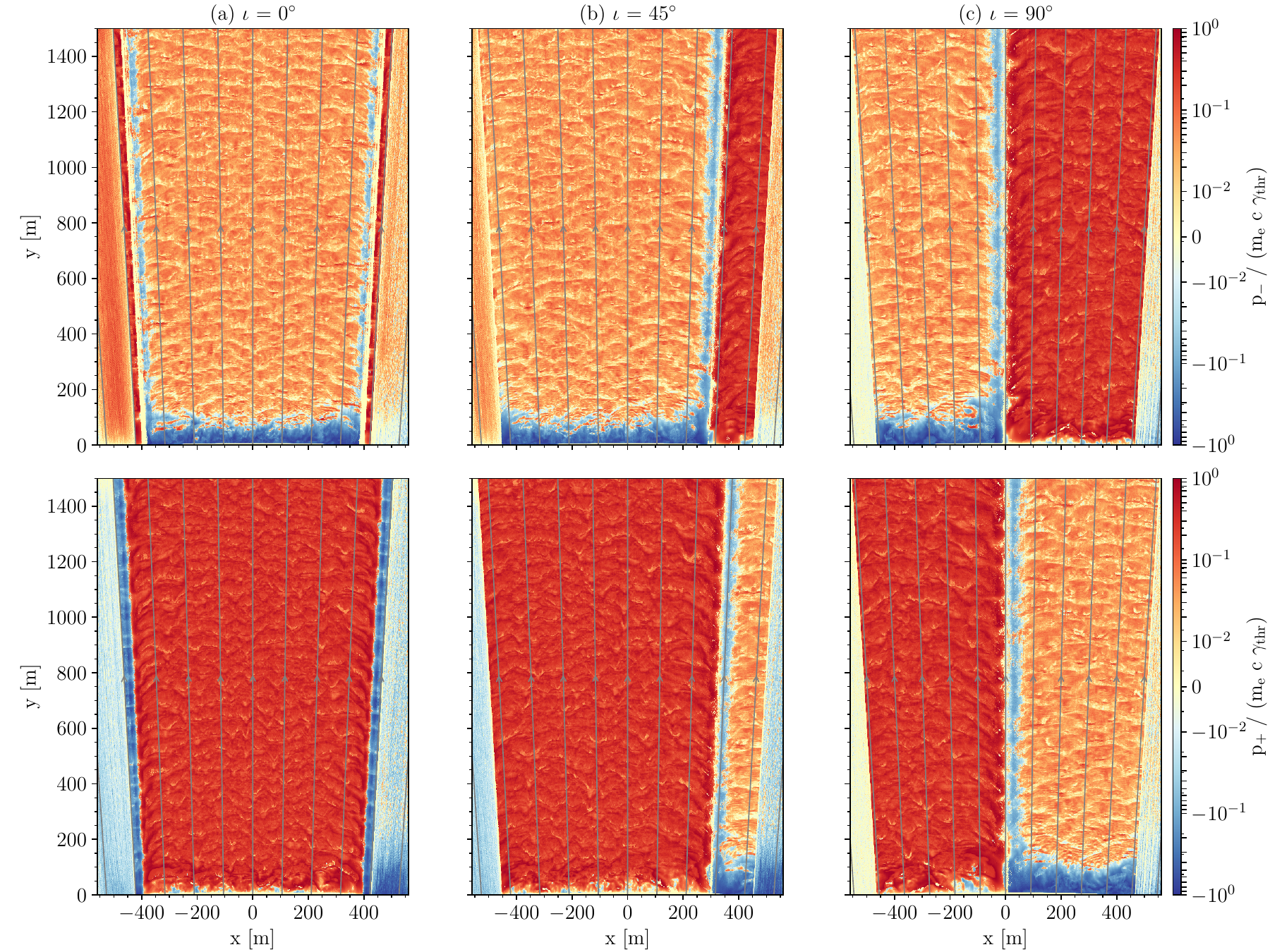}
  \captionof{figure}{Same as Fig.~\ref{fig2} but for the electron (top row) and positron (bottom row) bulk velocities.}\label{fig5}
\end{minipage}

}

\end{appendix}

\end{document}